\documentclass[a4,11pt]{article}
\usepackage{graphicx} 
\usepackage{amsmath}
\usepackage{amsfonts}
\usepackage{latexsym}
\usepackage{amssymb}
\makeatletter
 
  \@addtoreset{equation}{section}
 \makeatother

\begin{document}
\title{Derivative Pricing under Asymmetric and Imperfect Collateralization and CVA~\footnote{
This research is supported by CARF (Center for Advanced Research in Finance) and 
the global COE program ``The research and training center for new development in mathematics.''
All the contents expressed in this research are solely those of the authors and do not represent any views or 
opinions of any institutions. 
The authors are not responsible or liable in any manner for any losses and/or damages caused by the use of any contents in this research.
}
%\\ \Large{\it{--Implications of asymmetric CSA and suboptimal strategies--}}
}

\author{Masaaki Fujii\footnote{Graduate School of Economics, The University of Tokyo},
Akihiko Takahashi\footnote{Graduate School of Economics, The University of Tokyo}
}
%\begin{center}
\date{
First version: November 30, 2010\\
Current version: December 15, 2011 %\today
}
%\end{center}
\maketitle

%%%%%%    TEXT START    %%%%%%

%%%%%%      Macros      %%%%%%
%nakamacro.tex(H120522;0730)
%\documentstyle[11pt]{article}
%\setlength{\textwidth}{10.5in}
%\setlength{\oddsidemargin}{0in}
%\setlength{\topmargin}{-0.52in}
%\setlength{\textheight}{9.0in}
%\setlength{\footskip}{0.7in}

\newtheorem{definition}{Definition}
\newtheorem{assumption}{$[$ A}
\newtheorem{condition}{$[$ C}
\newtheorem{lemma}{Lemma}
\newtheorem{proposition}{Proposition}
\newtheorem{theorem}{Theorem}
\newtheorem{remark}{Remark}
\newtheorem{example}{Example}
\newtheorem{corollary}{Corollary}
%--------------------------------------------------------------------------
%BOLD FACES
\def\n{{\bf n}}
\def\A{{\bf A}}
\def\B{{\bf B}}
\def\C{{\bf C}}
\def\D{{\bf D}}
\def\E{{\bf E}}
\def\F{{\bf F}}
\def\G{{\bf G}}
\def\H{{\bf H}}
\def\I{{\bf I}}
\def\J{{\bf J}}
\def\K{{\bf K}}
\def\L{{\bf L}}
\def\M{{\bf M}}
\def\N{{\bf N}}
\def\O{{\bf O}}
\def\P{{\bf P}}
\def\Q{{\bf Q}}
\def\R{{\bf R}}
\def\S{{\bf S}}
\def\T{{\bf T}}
\def\U{{\bf U}}
\def\V{{\bf V}}
\def\W{{\bf W}}
\def\X{{\bf X}}
\def\Y{{\bf Y}}
\def\Z{{\bf Z}}
\def\cala{{\cal A}}
\def\calb{{\cal B}}
\def\calc{{\cal C}}
\def\cald{{\cal D}}
\def\cale{{\cal E}}
\def\calf{{\cal F}}
\def\calg{{\cal G}}
\def\calh{{\cal H}}
\def\cali{{\cal I}}
\def\calj{{\cal J}}
\def\calk{{\cal K}}
\def\call{{\cal L}}
\def\calm{{\cal M}}
\def\caln{{\cal N}}
\def\calo{{\cal O}}
\def\calp{{\cal P}}
\def\calq{{\cal Q}}
\def\calr{{\cal R}}
\def\cals{{\cal S}}
\def\calt{{\cal T}}
\def\calu{{\cal U}}
\def\calv{{\cal V}}
\def\calw{{\cal W}}
\def\calx{{\cal X}}
\def\caly{{\cal Y}}
\def\calz{{\cal Z}}
%
%YOKUTUKAUMONO
\def\sskip{\hspace{0.5cm}}
\def\simleq{ \raisebox{-.7ex}{\em $\stackrel{{\textstyle <}}{\sim}$} }
\def\leqsim{ \raisebox{-.7ex}{\em $\stackrel{{\textstyle <}}{\sim}$} }
\def\ep{\epsilon}
\def\half{\frac{1}{2}}
\def\iku{\rightarrow}
\def\Iku{\Rightarrow}
\def\ikup{\rightarrow^{p}}
\def\inclusion{\hookrightarrow}
\def\cadlag{c\`adl\`ag\ }
\def\up{\uparrow}
\def\down{\downarrow}
\def\doti{\Leftrightarrow}
\def\douti{\Leftrightarrow}
\def\dochi{\Leftrightarrow}
\def\douchi{\Leftrightarrow}%
%KAIGYOU,ARRAY
\def\yy{\\ && \nonumber \\}
\def\y{\vspace*{3mm}\\}
\def\nn{\nonumber}
\def\be{\begin{equation}}
\def\ee{\end{equation}}
\def\bea{\begin{eqnarray}}
\def\eea{\end{eqnarray}}
\def\beas{\begin{eqnarray*}}
\def\eeas{\end{eqnarray*}}
%
%KONO RONBUN DE TUKAU MONO
\def\hd{\hat{D}}
\def\hv{\hat{V}}
\def\hsd{{\hat{d}}}
\def\hx{\hat{X}}
\def\hsx{\hat{x}}
\def\bsx{\bar{x}}
\def\bsd{{\bar{d}}}
\def\bx{\bar{X}}
\def\ba{\bar{A}}
\def\bb{\bar{B}}
\def\bc{\bar{C}}
\def\bv{\bar{V}}
\def\balpha{\bar{\alpha}}
\def\bbalpha{\bar{\bar{\alpha}}}
\def\combi{\l(\begin{array}{c}\alpha\\ \beta \end{array}\r)}
\def\f{^{(1)}}
\def\s{^{(2)}}
\def\ss{^{(2)*}}
\def\l{\left}
\def\r{\right}
\def\a{\alpha}
\def\b{\beta}
\def\L{\Lambda}
%上に定義されたコマンドは数式モ−ドで用いる。
%--------------------------------------------------

\def\E{{\bf E}}
\def\P{{\bf P}}
\def\Q{{\bf Q}}
\def\R{{\bf R}}

\def\calf{{\cal F}}
\def\calp{{\cal P}}
\def\calq{{\cal Q}}

\def\ep{\epsilon}

\def\yy{\\ && \nonumber \\}
\def\y{\vspace*{3mm}\\}
\def\nn{\nonumber}
\def\be{\begin{equation}}
\def\ee{\end{equation}}
\def\bea{\begin{eqnarray}}
\def\eea{\end{eqnarray}}
\def\beas{\begin{eqnarray*}}
\def\eeas{\end{eqnarray*}}
\def\l{\left}
\def\r{\right}
\vspace{10mm}

%%%%%%%%%%%%%%%%%%%%%%%%%%%%%%$$$
\begin{abstract}
The importance of collateralization through the change of funding cost 
is now well recognized among practitioners. 
In this article, we have extended the previous studies of collateralized derivative pricing 
to more generic situation, that is asymmetric and imperfect collateralization with the associated
counter party credit risk. By introducing the collateral coverage ratio, our framework can handle these
issues in an unified manner. Although the resultant pricing formula becomes non-linear FBSDE
and cannot be solve exactly, the fist order approximation is provided using 
Gateaux derivative. We have shown that it allows us to decompose the price of generic contract into three parts:
market benchmark, bilateral credit value adjustment (CVA), and the collateral cost adjustment (CCA) independent from the credit risk. We have studied each term closely, and demonstrated the significant impact of asymmetric collateralization 
through CCA using the numerical examples.

\end{abstract}
\vspace{17mm}
%%%%%%%%%%%%%%%%%%%%%%%%%%%%%%%%%$

\newpage
%%%%%%%%%%%%%%%%%%%%%%%%%%%%%%%%%
\section{Introduction}
%%%%%%%%%%%%%%%%%%%%%%%%%%%%%%%%%
In the last decade, collateralization has experienced dramatic increase in the derivative market.
According to the ISDA survey~\cite{ISDA}, the percentage of trade volume subject to collateral agreements in  the OTC (over-the-counter) 
market has now become $70\%$, which was merely $30\%$ in 2003. If we focus on 
large broker-dealers and the fixed income market, the coverage goes up even higher to $84\%$.
Stringent collateral management is also a crucial issue for successful installation of 
CCP (central clearing parties).

Despite its long history in the financial market as well as its critical role in the 
risk management, it is only after the explosion of Libor-OIS spread
following the collapse of Lehman Brothers that the effects of collateralization on 
derivative pricing have started to gather strong attention among practitioners.
In most of the existing literatures, collateral cost has been neglected, and only its
reduction of counterparty exposure have been considered.
The work of Johannes \& Sundaresan (2007)~\cite{collateralized_swap} was the first 
focusing on the cost of collateral, which studied 
its effects on swap rates based on empirical analysis.
As a more recent work, Piterbarg (2010)~\cite{Piterbarg} discussed the general option pricing 
using the similar formula to take the collateral cost into account.

In a series of works of Fujii, Shimada \& Takahashi (2009)~\cite{multiple_curves, dynamic_basis}
and Fujii \& Takahashi (2010,2011)~\cite{collateral_termstructures, collateral_choice}, 
modeling of interest rate term structures 
under collateralization has been studied, where cash collateral is assumed to be posted 
continuously and hence the remaining counterparty credit risk is  negligibly small.
In these works, it was found that there exists a direct link between the cost of collateral and 
CCS (cross currency swap) spreads.
In fact, one cannot neglect the cost of collateral to make the whole system consistent with 
CCS markets, or equivalently with FX forwards.
Making use of this relation, we have also shown the significance of 
a "cheapest-to-deliver" (CTD) option implicitly embedded in a collateral agreement in 
Fujii \& Takahashi (2011)~\cite{collateral_choice}.

In this paper, we have extended the existing works to handle 
asymmetric as well as imperfect collateralization in the unified credit-risk modeling framework.
Asymmetric collateralization arises when CSA (credit support annex) treats the two contracting parties
asymmetrically, such as different collateral thresholds and unilateral collateralization.
Even if the adopted CSA is symmetric, asymmetric collateralization may arise due to the different level
of sophistication in collateral management of the two parties. For example, even if CSA allows the same set of 
eligible currencies for the both parties, if one of them does not have appropriate system or easy access to the 
relevant foreign currencies, it cannot fully exercise the allowed optionality of choosing the cheapest currency
to post. We have shown interesting numerical examples using OIS (overnight index swap)
and CCS.

By introducing the collateral coverage ratio, our pricing framework allows to include
under- as well as over-collateralization where there remains counter party credit risk.
In general, we see the pricing formula becomes non-linear FBSDEs since the amount of collateral and also the 
counter-party exposure depend on the value of portfolio at every time in the future, 
which itself is affected by the collateral cost and default payoff in turn.
We have adopted Gateaux derivative to obtain the first order approximation, 
as in the work of Duffie \& Huang (1996)~\cite{Duffie},
and found that it is possible to decompose the price of portfolio into the following parts:\\
~(1) Clean price representing the value under the perfect and symmetric collateralization.\\
~(2) Bilateral CVA (or CVA/DVA).\\
~(3) CCA (collateral cost adjustment) representing the change of collateral cost due to the 
deviation from the perfect symmetric collateralization but independent from credit risk.
\\\\
We will see that the {\it clean price} under the perfect and symmetric collateralization has the desirable features 
as the market benchmark. It retains the additivity and hence one can evaluate each trade (and cash flow) 
separately. In fact, we see that the clean price formula is consistent with those obtained
in our previous works~\cite{multiple_curves, collateral_choice}, which is becoming the market standard
as the OIS discounting.
In other words, our pricing method allows one to decompose the value of a generic contract into the 
market benchmark and the remaining corrections which are credit risk and collateral cost.

The organization of the paper is as follows: In Sec.~\ref{formulation}, we discuss the 
generic formulation. The first order approximation using Gateaux derivative is explained in Sec.~\ref{sec_decomposition}. 
Secs.~\ref{perfect_collateralization} and \ref{instruments}
are devoted to explain the case of perfect collateralization and 
Sec.~\ref{numerical} provides numerical examples to demonstrate the effects of asymmetric collateralization.
Sec.~\ref{asymmetric_implications} discusses the implications for  the behavior of 
financial firms induced by these effects. Sec.~\ref{imp_collateral} treats the imperfect collateralization and CVA, 
which shows the importance of the interplay between the collateral funding cost and other variables. We finally conclude 
in Sec.~\ref{conclusion}. The following Appendix contains technical details and proofs omitted in the main text.

%%%%%%%%%%%%%%%%%%%%%%%%%%%%%%%%%%%%%%%%%%%%%%%%%%%%%%%%%%%%%%%%%%%%%%%%%%%%
\section{Generic Formulation}
\label{formulation}
%%%%%%%%%%%%%%%%%%%%%%%%%%%%%%%%%%%%%%%%%%%%%%%%%%%%%%%%%%%%%%%%%%%%%%%%%%%%%
In this section, we consider the generic pricing formula.
As an extension from the previous works, we allow asymmetric and/or 
imperfect collateralization with bilateral default risk.
We basically follow the setup in Duffie \& Huang (1996)~\cite{Duffie} 
and extend it so that we can deal with existence of collateral and its cost explicitly.
The approximate pricing formulas that allow simple analytic treatment
are derived by Gateaux derivatives. See other related developments in, for examples,
\cite{Jeanblanc, Brigo_Capponi, lipton_sepp, Bielecki_book, oxfordbook} 
and references therein.

%%%%%%%%%%%%%%%%%%%%%%%%%%%%%%%%%%%%%%%%%%%%%%%%
\subsection{Fundamental Pricing Formula}
%%%%%%%%%%%%%%%%%%%%%%%%%%%%%%%%%%%%%%%%%%%%%%%%
\subsubsection{Setup}\label{setup}
%%%%%%%%%%%%%%%%%%%%%%%%%%%%%%%%%%%%%%%%%%%%%%%%
We consider a filtered probability space $(\Omega, \calf, \mathbb{F}, Q)$,
where $\mathbb{F}=\{\calf_t:t\geq 0\}$ is sub-$\sigma$-algebra of $\calf$ satisfying 
the {\it usual} conditions. Here, $Q$ is the spot martingale measure, 
where the money market account is being used as the numeraire.
We consider two counterparties, which are denoted by party $1$ and party $2$.
We model the stochastic default time of party $i$ $(i\in\{1, 2\})$ as an 
$\mathbb{F}$-stopping time $\tau^i \in [0,\infty]$, which are 
assumed to be totally inaccessible.
We introduce, for each $i$, the default indicator function,
$H_t^{i}=\bold{1}_{\{\tau^i\leq t\}}$, a stochastic process that is
equal to one if party $i$ has defaulted, and zero otherwise.
The default time of any financial contract between the two parties is 
defined as $\tau=\tau^1 \wedge \tau^2$, the minimum of $\tau^1$ and $\tau^2$.
The corresponding default indicator function of the contract is denoted by
$H_t=\bold{1}_{\{\tau\leq t\}}$. The Doob-Meyer theorem implies the existence of the unique decomposition 
as $H^i=A^i + M^i$, where $A^i$ is a predictable and right-continuous (it is continuous indeed, 
since we assume total inaccessibility of default time), 
increasing process with $A^i_0=0$, and $M^i$ is a $Q$-martingale.
In the following, we also assume the absolute continuity of $A^i$ and the 
existence of progressively measurable non-negative process $h^i$,
usually called the hazard rate of counterparty $i$, such that
\be
A_t^i =\int_0^t h_s^i \bold{1}_{\{\tau^i>s\}}ds, \qquad t\geq 0~.
\ee
For simplicity we also assume that there is no simultaneous default with positive probability and 
hence the hazard rate for $H_t$ is given by $h_t=h_t^1+h_t^2$ on the set of $\{\tau>t\}$.

%%%%%%%%%%%%%%%%%%%%%%%%%%%%%%%%%%%%%%%%%%%%%
\subsubsection{Collateralization}
\label{margining}
%%%%%%%%%%%%%%%%%%%%%%%%%%%%%%%%%%%%%%%%%%%%%
We assume collateralization by cash which works in the following way:
{\it if the party $i~(\in\{1,2\})$ has negative mark-to-market, it has to post cash 
collateral~\footnote{According to the ISDA survey~\cite{ISDA}, more than $80\%$ of collateral 
being used is cash. If there is a liquid repo or security-lending market, we may also carry out 
similar formulation with proper adjustments of its funding cost.} to the 
counter party $j~(\neq i)$, where the coverage ratio of the exposure is denoted by $\delta_t^i\in\mathbb{R}_{+}$.
We assume the margin call and settlement occur instantly.
Party $j$ is then a collateral receiver and has to pay collateral rate $c_t^{i}$ on 
the posted amount of collateral, which is $\delta_t^{i}\times (|$mark-to-market$|)$, to the party $i$. 
This is done continuously until the end of the contract.
A common practice in the market is to set $c_t^{i}$ as the 
time-$t$ value of overnight (ON) rate of the collateral currency used by the party $i$.}

We emphasize that it is crucially important to distinguish the ON rate $c^i$ from 
the theoretical risk-free rate of the same currency $r^i$, where both of them are progressively 
measurable. The distinction is necessarily for 
the unified treatment of different collaterals and for the consistency with 
cross currency basis spreads, or equivalently FX forwards in the 
market~(See, Sec.~\ref{MtMCCOIS} and Ref.~\cite{collateral_choice} for details.).

We consider the assumption of continuous collateralization is a reasonable proxy of the current market 
where daily (even intra-day) margin call is becoming popular.
Although we assume continuous collateralization, we include the under- as well as over-collateralization
in which we have $\delta_t^i<1$ and $\delta_t^i>1$, respectively.
It may look slightly odd at first sight to include $\delta_t^i\neq 1$ cases under the continuous assumption,
we think that allowing it makes the model more realistic considering the 
possible price dispute between the relevant parties, which is particularly the case for exotic derivatives.
Because of the model uncertainty, the price reconciliation is usually done 
in ad-hoc way, say taking an average of each party's quote. As a result, even after the 
each margin settlement, there always remains sizable discrepancy between the collateral value and the
model implied fair value of the portfolio. Furthermore, in order to prepare for the rapid 
change of the collateral value, over-collateralization with sizable haircut is also frequently 
observed in the market.

Under these assumptions, the remaining credit exposure of the party $i$ to the party $j$  at time $t$ is given by 
\be
\max(1-\delta_t^j,0)\max(V^{i}_t,0)+\max(\delta_t^i-1,0)\max(-V_t^{i},0)~,\nonumber
\ee
where $V_t^{i}$ denotes the mark-to-market value of the contract from the view point of party $i$.
The second term corresponds to the over-collateralization, where the party $i$ can 
only recover the fraction of overly posted collateral when party $j$ defaults.
We denote the recovery rate of the party $j$, when it defaults at time $t$, 
by the progressively measurable process $R_t^j\in[0,1]$.
Thus, the recovery value that the party $i$ receives can be written as
\be
R_t^{j}\left(\max(1-\delta_t^j,0)\max(V^{i}_t,0)+\max(\delta_t^i-1,0)\max(-V_t^{i},0)\right)~.
\ee

As for notations, we will use a bracket "$(~)$" when we specify type of currency, 
such as $r_t^{(i)}$ and $c_t^{(i)}$, the risk-free and the collateral rates of currency $(i)$,  
in order to distinguish it from that of counter party.
We also denote a spot FX at time $t$ by $f_{x}^{(i,j)}(t)$ that is the price of a unit amount of currency $(j)$ in 
terms of currency $(i)$. We assume all the technical conditions for integrability are satisfied 
throughout this paper.

%%%%%%%%%%%%%%%%%%%%%%%%%%%%%%%%%%%%%%%%%%%%%%%%%%%%%%%%%%
\subsubsection{Pricing Formula}
%%%%%%%%%%%%%%%%%%%%%%%%%%%%%%%%%%%%%%%%%%%%%%%%%%%%%%%%%%
We consider the ex-dividend price at time $t$ of a generic financial contract made
between the party $1$ and $2$, whose maturity is set as $T~(>t)$. 
We consider the valuation from the view point of party $1$,
and define the cumulative dividend $D_t$ that is the total 
receipt from party $2$ subtracted by the total payment from party $1$.
We denote the contract value as $S_t$ and define $S_t=0$ for $\tau\leq t$.
See Ref.\cite{Duffie} for the technical details about the regularity conditions which 
guarantee the existence and uniqueness of $S_t$.

Under these assumptions and the setup give in Secs.~\ref{setup} and \ref{margining}, one obtains
\bea
S_t&=&\beta_t E^Q\left[\int_{]t,T]}\beta_u^{-1}\bold{1}_{\{\tau>u\}}
\Bigl\{ dD_u + \left(y_u^{1}\delta_u^1\bold{1}_{\{S_{u}<0\}}+
y_u^2 \delta_u^2 \bold{1}_{\{S_{u}\geq 0\}}\right)S_{u}du\Bigr\}\right. \nonumber\\
&&\qquad\quad+\left.\left.\int_{]t,T]}\beta_u^{-1}\bold{1}_{\{\tau\geq u\}}\Bigl(
Z^1(u,S_{u-})dH_u^1+Z^2(u,S_{u-})dH_u^2\Bigr)\right|\calf_t\right],
\label{generic_pricing1}
\eea
on the set of $\{\tau>t\}$. Here, $y^i=r^i-c^i$ denotes a spread 
between the risk-free and collateral rates of the currency used by party $i$,
which represents the instantaneous return from the collateral being posted, i.e. 
it earns $r^i$ but subtracted by $c^i$ as the payment to the collateral payer.
Here, we have used the risk-free rate as the effective investment return or borrowing cost of cash after adjusting all the market and credit risks.
$\beta_t=\exp\left(\int_0^t r_u du\right)$ is a money market account for the currency 
on which $S_t$ is defined. $Z^i$ is the recovery payment from the view point of the party $1$ at the 
time of default of party $i~(\in\{1,2\})$:
\bea
&&\hspace{-7mm}Z^1(t,v)=\Bigl(1-(1-R_t^1)(1-\delta_t^1)^+\Bigr) v\bold{1}_{\{v<0\}}+
\Bigl(1+(1-R_t^1)(\delta_t^2-1)^+\Bigr)v\bold{1}_{\{v\geq 0\}}\\
&&\hspace{-7mm}Z^2(t,v)=\Bigl(1-(1-R_t^2)(1-\delta_t^2)^+\Bigr) v\bold{1}_{\{v\geq 0\}}
+\Bigl(1+(1-R_t^2)(\delta_t^1-1)^+\Bigr)v\bold{1}_{\{v<0\}}~,
\eea
where $X^+$ denotes $\max(X,0)$.
Note that the above definition is consistent with the setup in Sec.\ref{margining}.
The surviving party loses money if the received collateral from the defaulted party is not 
enough or if the posted collateral to the defaulted party exceeds the fair contract value.

Eq.~(\ref{generic_pricing1}) contains the indicator function $H^i$ within the expectation and is not useful
for valuation. Thus, as usually done in the credit modeling, we try to eliminate the indicators
from the expectation. Even in the presence of the collateral, it is in fact
possible to prove the following proposition in completely parallel fashion with the one given in \cite{Duffie}:
%%%%%%%%%%%%%%%%%%%%%%%%%%%%%%%%%%%%%%%%%%%%%%%%%%%%%%%%%%%%%%%%%%%%%%%%%
\begin{proposition} \label{prop-1}
Suppose a generic financial contract between the party $1$ and $2$, of which cumulative dividend at time $t$ 
is denoted by $D_t$ from the view point of the party $1$.
Assume that the contract is continuously collateralized by cash where the coverage ratio 
of the party $i~(\in\{1,2\})$'s exposure is denoted by $\delta_t^i\in \mathbb{R}_+$. The collateral receiver $j$
has to pay the  collateral rate denoted by $c_t^i$ on the amount of collateral posted by party $i$,
which is not necessarily equal to the risk-free rate of the same  currency, $r_t^i$.
The fractional recovery rate $R_t^i\in[0,1]$ is assumed 
for the under- as well as  over-collateralized exposure.
For the both parties, totally inaccessible default is assumed, and the  
hazard rate process of party $i$ is denoted by $h_t^i$. We assume there is no 
simultaneous default of the party $1$ and $2$, almost surely. 

Then, conditioned on no-default $(\tau>t)$, the contract value $S_t$ given in Eq.~(\ref{generic_pricing1})
is represented by the pre-default value $V_t$ satisfying 
$S_t=V_t\bold{1}_{\{\tau>t\}}$
\be\label{eqV}
V_t=E^Q\left[\left.\int_{]t,T]}\exp\left(-\int_t^s \bigl(r_u-\mu(u,V_u)\bigr)du\right)dD_s\right|\calf_t\right]~,
\quad t\leq T
\ee
where
\bea
\mu(t,v)&=&\Bigl( y_t^1\delta_t^1-(1-R_t^1)(1-\delta_t^1)^+h_t^1+(1-R_t^2)(\delta_t^1-1)^+h_t^2\Bigr)
\bold{1}_{\{v<0\}}\nonumber\\
&+&\Bigl(y_t^2\delta_t^2-(1-R_t^2)(1-\delta_t^2)^+h_t^2+(1-R_t^1)(\delta_t^2-1)^+h_t^1\bigr)
\bold{1}_{\{v\geq 0\}}
\eea
if the jump of $V$ at the time of default$~(=\tau)$ is zero almost surely.
\end{proposition}
%%%%%%%%%%%%%%%%%%%%%%%%%%%%%%%%%%%%%%%%%%%%%%%%%%%%%%%%%%%%%%%%%%%%%%%%%%%
See Appendix~\ref{proof-1} for the proof. 
\\\\
Naively speaking, by focusing on the pre-default world $(\tau>t)$,
we have replaced all the indicator functions in (\ref{generic_pricing1}) with the corresponding intensities 
(or hazard rates), which then leads to the expression of $V$ in (\ref{eqV}) by straightforward integration.
However, precisely speaking, we have to check if the two quantities $S_t$ and $V_t\bold{1}_{\{\tau>t\}}$  actually 
follow the equivalent SDEs up to the default time. This point is actually confirmed in the proof.
Note that the no-jump condition $\Delta V_{\tau}=0$ is not crucial. Since we are only interested in 
pre-default value, we can replace the hazard rates as those conditioned on no-default, which then
easily recovers $\Delta V_{\tau}=0$. See the remark in Appendix~\ref{proof-1} for the details.

%%%%%%%%%%%%%%%%%%%%%%%%%%%%%%%%%%%%%%%%%%%%%%%%%%%%%%%%%%%%%%%%%%%%%%%%%%%%%%
\section{Decomposing the Pre-default Value}
\label{sec_decomposition}
%%%%%%%%%%%%%%%%%%%%%%%%%%%%%%%%%%%%%%%%%%%%%%%%%%%%%%%%%%%%%%%%%%%%%%%%%%%%%%%
From Proposition~\ref{prop-1}, one sees the effective discounting rate becomes non-linear 
due to the correction term $\mu(t,v)$:
\bea
\mu(t,v)=\tilde{y}^{1}_t\bold{1}_{\{v<0\}}+\tilde{y}^2_t\bold{1}_{\{v\geq 0\}}
\eea 
where
\bea
\tilde{y}^{i}_t=\delta^{i}_t y_t^i-(1-R_t^i)(1-\delta_t^i)^+h_t^i+(1-R_t^j)(\delta_t^i-1)^+h_t^j
\eea
for $i\in\{1,2\}$ and $j\neq i$. As one can see, due to the presence of the dependence on $v$
through the indicator, the pricing formula of Eq.~(\ref{eqV}) together with other market variables 
form a system of non-linear FBSDE.
Since it is impossible to solve in general, we need to consider some 
approximation procedures. 

%%%%%%%%%%%%%%%%%%%%%%%%%%%%%%%%%%%%%%%%%%%
\subsection{Perfect and Symmetric Collateralization}
%%%%%%%%%%%%%%%%%%%%%%%%%%%%%%%%%%%%%%%%%%%
Since non-linearity appears only through the indicators, we can recover the linearity if 
\be
\tilde{y}^1_s=\tilde{y}^2_s  \quad \mbox{Lebesgue-a.e.}~~ s\in[t,T]~~\mbox{a.s.}
\ee
This situation naturally arises
when the collateralization is perfect and symmetric, or $(\delta^1=\delta^2=1)$ and $(y^1= y^2=\overline{y})$.
In fact, in this case, the pricing formula (\ref{eqV}) becomes
\bea
\overline{V}_t=E^Q\left[\left.\int_{]t,T]}\exp\left(-\int_t^s \bigl(r_u-\overline{y}_u\bigr)du\right)dD_s\right|\calf_t\right]~.
\label{eqVbar}
\eea
Note that we have recovered the additivity here: The portfolio value can be obtained 
by adding that of each trade, which is calculable separately. This is the most important feature as the market benchmark 
since otherwise a trade price depends on the whole portfolio to the specific counter party and there would be no
price transparency. As we will see shortly, this actually corresponds to the collateral rate discounting~\cite{multiple_curves, Piterbarg}, which is becoming the market standard~\cite{KPMG}.
%%%%%%%%%%%%%%%%%%%%%%%%%%%%%%%%%%%%%%%%%%%
\subsection{Generic Situations}
\label{sec_generic}
%%%%%%%%%%%%%%%%%%%%%%%%%%%%%%%%%%%%%%%%%%%
Now, let us consider more generic situations where the collateralization is imperfect $(\delta^i\neq 1)$
and/or asymmetric $(y^1\neq y^2)$. In order to evaluate Eq.~(\ref{eqV}) approximately, we consider 
expanding $V$ around the previously obtained $\overline{V}$ in the first order of Taylor expansion. It means that 
we try to express the contract value by the market benchmark  and the
remaining corrections.

Firstly, let us express the correction to the discounting rate $\mu(t,v)$ around 
the collateral cost at the benchmark point $\overline{y}$:
\bea
\mu(t,v)=\overline{y}_t+\Delta \tilde{y}^1_t\bold{1}_{\{v<0\}}+
\Delta \tilde{y}^2_t\bold{1}_{\{v\geq 0\}}
\eea
where
\bea
\Delta \tilde{y}^i_t&=&\tilde{y}^i_t-\overline{y}_t\nn \\
&=&\Bigl\{(\delta_t^i y_t^i-\overline{y}_t)-(1-R_t^i)(1-\delta_t^i)^+h_t^i
+(1-R_t^j)(\delta_t^i-1)^+h_t^j\Bigr\}
\eea
for $i\in\{1,2\}$ and $j\neq i$. The first order effect of $\Delta\tilde{y}$ 
is given by Gateaux derivative $\nabla V$, which is a sort of gradient by considering $V$ as a functional of $\tilde{y}$:
\bea
\lim_{\ep \downarrow 0}\sup_t \left|
\nabla{V}(\overline{y},\tilde{y})-\frac{V_t\Bigl(\overline{y}+\epsilon \bigl(\Delta \tilde{y}^1_t\bold{1}_{\{v<0\}}+
\Delta \tilde{y}^2_t\bold{1}_{\{v\geq 0\}}\bigr)\Bigr)-V_t(\overline{y})}{\epsilon}\right|~.
\eea
Notice that $V_t(\overline{y})$ is actually $\overline{V}_t$ in Eq.~(\ref{eqVbar}).
Following the method explained in Duffie \& Skiadas (1994)~\cite{Duffie-Skiadas} and Duffie \& Huang (1996)~\cite{Duffie} , we can derive
\bea
\nabla V(\overline{y},\tilde{y})=E^Q\left[\left.\int_t^T e^{-\int_t^s (r_u-\overline{y}_u)du}\overline{V}_s
\Bigl(\Delta \tilde{y}^1_s\bold{1}_{\{\overline{V}_s<0\}}+
\Delta \tilde{y}_s^2 \bold{1}_{\{\overline{V}_s\geq 0\}}\Bigr)ds\right|\calf_t\right]~.
\label{eq_gateaux}
\eea

Substituting the contents of $\Delta\tilde{y}$ into (\ref{eq_gateaux}), 
the above Gateaux derivative can be further decomposed into two parts, one of which is  
the collateral cost adjustment independent from the credit risk
\bea
{\rm CCA}_t=E^Q\left[\left. \int_t^T e^{-\int_t^s (r_u-\overline{y}_u)du}
\Bigl( -(\delta_s^1 y_s^1-\overline{y}_s)\bigl[-\overline{V}_s\bigr]^+
+(\delta_s^2 y_s^2-\overline{y}_s)\bigl[\overline{V}_s\bigr]^+\Bigr)ds\right|\calf_t\right]
\label{generic_cca}
\eea
and the other is the well-known bilateral CVA terms
\bea
{\rm CVA}_t&=&E^Q\left[\left. \int_t^T e^{-\int_t^s (r_u-\overline{y}_u)du}
h_s^1(1-R_s^1)\Bigl\{(1-\delta_s^1)^+\bigl[-\overline{V}_s]^+
+(\delta_s^2-1)^+\bigl[\overline{V}_s\bigr]^+\Bigr\}ds\right|\calf_t\right] \nn \\
&-&E^Q\left[\left. \int_t^T e^{-\int_t^s (r_u-\overline{y}_u)du}h_s^2(1-R_s^2)
\Bigl\{(1-\delta_s^2)^+\bigl[\overline{V}_s\bigr]^++
(\delta_s^1-1)^+\bigl[-\overline{V}_s\bigr]^+\Bigr\}ds\right|\calf_t\right]~.\nn\\
\label{generic_cva}
\eea
Using these terms, the original pre-default value of Eq.~(\ref{eqV}) can be approximated
in the first order of $\Delta\tilde{y}$ as~\footnote{
Recently, we have developed the methodology to carry out higher order approximation for 
non-linear FBSDEs~\cite{nl-FBSDE}.}
\bea
V_t=\overline{V}_t+{\rm CCA}_t+{\rm CVA}_t +o(\Delta \tilde{y}^1,\Delta\tilde{y}^2)~.
\eea
In the following sections, we will study each term more closely.
We basically treat the perfect collateralization $(\delta^1=\delta^2=0)$ until
Sec.~\ref{imp_collateral},  where we will study CVA.

%%%%%%%%%%%%%%%%%%%%%%%%%%%%%%%%%%%%%%%%%%%%%%%
\section{Perfect Collateralization}
\label{perfect_collateralization}
%%%%%%%%%%%%%%%%%%%%%%%%%%%%%%%%%%%%%%%%%%%%%%%%
In the following section, we first deal with the perfectly collateralized situation where ${\rm CVA}=0$.
We check the several important examples, which will also be used to 
price some fundamental instruments used in numeral examples in the next section.

%%%%%%%%%%%%%%%%%%%%%%%%%%%%%%%%%%%%%%%%%%%%%%%%
\subsection{Symmetric Collateralization}
\label{symmetric}
%%%%%%%%%%%%%%%%%%%%%%%%%%%%%%%%%%%%%%%%%%%%%%%%
Let us first consider the market benchmark, or the perfect and symmetric collateralization.
In this case we have $\delta^1=\delta^2=1$ and $y^1=y^2=\overline{y}$, and hence ${\rm CCA}={\rm CVA}=0$.
One can easily confirm that all the following results are consistent with those given in 
Refs.~\cite{multiple_curves, dynamic_basis,collateral_choice,collateral_termstructures}.
\\
\\
$\bold{Case~1}$:
Situation where 
both parties use the same collateral currency "$(i)$", which is the same as the deal currency.
Since we have $\overline{y}=y^{(i)}=r^{(i)}-c^{(i)}$ in this case, 
the pre-default value of the contract in terms of currency $(i)$ is given by
\be
V_t=E^{Q^{(i)}}\left[\left.\int_{]t,T]}\exp\left(-\int_t^s c_u^{(i)}du\right)dD_s\right|\calf_t\right]~,
\label{perfect_single}
\ee
where $Q^{(i)}$ is the spot-martingale measure of currency $(i)$. In this case, we can use the collateral rate
to discount the future cash flows as if it is the usual risk-free rate. 
Since the collateral rate of cash is the corresponding ON rate, 
this formula corresponds to the OIS discounting method, which is becoming the new standard in the 
market~\footnote{See the recent survey done by KPMG~\cite{KPMG}.}.
\\\\
$\bold{Case~2}$:
Situation where
both parties use the same collateral currency "$(k)$", which is the different from the deal currency "$(i)$".
In this case, $\overline{y}=y^{(k)}$ and the pre-default value of the contract in terms of currency $(i)$ is given by
\be
V_t=E^{Q^{(i)}}\left[\left. \int_{]t,T]}\exp\left(-\int_t^s \bigl(c_u^{(i)}+y_u^{(i,k)}
\bigr)du\right)dD_s\right|\calf_t\right]~,
\label{perfect_foreign}
\ee
where we have defined the {\it funding spread} between the currencies $(i)$ and $(k)$:
\bea
y^{(i,k)}_u &=& y_u^{(i)}-y_u^{(k)} \\
&=&\bigl(r_u^{(i)}-c_u^{(i)}\bigr)-\bigl(r_u^{(k)}-c_u^{(k)}\bigr)~.
\eea
This formula is particularly important for non-G5 currencies and multi-currency trades where it is quite common to use USD 
as the collateral currency $(k)$.  
There, the funding spread reflects the funding cost of USD relative to the deal currency $(i)$,
which is reflected in the corresponding cross currency basis swap. This point will be explained in Sec.~\ref{MtMCCOIS} in details. 
\\
\\
$\bold{Case~3}$:
Situation where the deal currency is $(i)$ and  the both parties have a common set of eligible collateral currencies denoted by $\calc$.
Note that only the collateral payer at each time has the right to choose the collateral to post.
In this case, $\overline{y}=\min_{k\in \calc}y^{(k)}$ and then we have
\be
V_t=E^{Q^{(i)}}\left[\left. \int_{]t,T]}\exp\left(-\int_t^s \bigl(c_u^{(i)}+\max_{k\in \calc}y_u^{(i,k)}
\bigr)du\right)dD_s\right|\calf_t\right]~
\ee
as the pre-default value of the contract in terms of currency $(i)$.

Notice that the collateral payer will choose the currency $(k)$ that minimizes the cost of collateral $(\min_{k\in \calc}y^{(k)})$,
which is equivalent to maximizing the effective discounting rate 
so that the payer achieves the smallest mark-to-market loss.
The optionality is crucially depends on the volatility of CCS and can be
numerically quite significant. See Ref.~\cite{collateral_choice} for details.

%%%%%%%%%%%%%%%%%%%%%%%%%%%%%%%%%%%%%%%%%%%%%%%%%%%%%
\subsection{Asymmetric Collateralization}
%%%%%%%%%%%%%%%%%%%%%%%%%%%%%%%%%%%%%%%%%%%%%%%%%%%%%
We now  consider the situation where ${\rm CVA}=0$ but there remains non-zero CCA due to the 
asymmetric collateralization $y^1\neq y^2$.
\\
\\
Suppose the situation where the trade is perfectly collateralized $(\delta^1=\delta^2=1)$ and 
the party $1$ can choose the optimal currency from the eligible set denoted by $\calc$ or $(y^1=\min_{k\in \calc}y^{(k)})$,
whereas the party $2$ can only use the single currency $(j)$ as collateral $(y^2=y^{(j)})$. 
Assume the deal currency is $(i)$. If we choose the center of expansion as $\overline{y}=y^{(j)}~(=y^2)$ and then 
we have
\be
{\rm CCA}_t=E^{Q^{(i)}}\left[\left.
\int_t^T \exp\left(-\int_t^s\bigl(c_u^{(i)}+y_u^{(i,j)}\bigr)du\right)\bigl[-\overline{V}_s]^+\max_{k\in \calc}y_s^{(j,k)}\right|\calf_t\right]~,
\ee
where
\be
\overline{V}_t=E^{Q^{(i)}}\left[\left.\int_{]t,T]}\exp\left(-\int_t^s \bigl(c_u^{(i)}+y_u^{(i,j)}\bigr)du\right)dD_s
\right|\calf_t\right]~.
\ee 
Since only the party $1$ has the cheapest-to-deliver optionality, the CCA adds the positive value to the 
contract~\footnote{Remember that we are calculating the contract value from the view point of the party $1$.}.
This value should be reflected in the contract price otherwise the party $2$ will suffer a loss by 
giving a free option to the party $1$.

Although we have assumed the asymmetric collateral agreement in the above example, similar situation 
can naturally arise even if the relevant CSA itself is symmetric. For example, even if the party $2$ has
the same eligible collateral set, if it lacks the easy access to the currencies involved,
then it ends up with the same situation. One can see that it is very dangerous to make a flexible collateral
agreement if there is no ability to fully exercise its embedded optionality, especially when the counter party is
more sophisticated in collateral management. We will give interesting numerical studies for the 
above example in Sec.~\ref{numerical}.

Another important example of asymmetric collateralization is the {\it one-way CSA} where collateralization is 
performed only unilaterally. This is actually common when sovereigns or central banks are involved as counter parties.
The detailed explanation of this situation is given in Sec.~\ref{imp_collateral} with default risk taken into 
account.

%%%%%%%%%%%%%%%%%%%%%%%%%%%%%%%%%%%%%%%%%%%%%%%%%%%%%%%%%%%%%%%%%%%%%%%%%%%%%%%%%%%%%
\section{Some Fundamental Instruments}
\label{instruments}
%%%%%%%%%%%%%%%%%%%%%%%%%%%%%%%%%%%%%%%%%%%%%%%%%%%%%%%%%%%%%%%%%%%%%%%%%%
In order to study the quantitative effects of collateralization, we firstly need to understand the clean price, 
i.e., $\overline{V}$. 
The details of generic term structure modeling under perfect collateralization are available  
in Refs~\cite{multiple_curves, dynamic_basis, collateral_choice}.
In this section, we just summarize some of the fundamental instruments required to understand 
the following numerical examples which demonstrates the impact of asymmetric collateralization.

%%%%%%%%%%%%%%%%%%%%%%%%%%%%%%%%%%%%%%%%%%%%%%%%%%%%%%%%%%%%%%%%%%%%%%%%%%%%
\subsection{Collateralized Zero Coupon Bond}
%%%%%%%%%%%%%%%%%%%%%%%%%%%%%%%%%%%%%%%%%%%%%%%%%%%%%%%%%%%%%%%%%%%%%%%%%%%%
We {\it define} the collateralized zero coupon bond of currency $(i)$ as
\bea
D^{(i)}(t,T)=E^{Q^{(i)}}\left[\left.e^{-\int_t^T c^{(i)}_sds}\bold{1}\right|\calf_t\right]~.
\eea
We call it "Bond" by analogy with conventional interest rate models 
but it simply represents the present value of the perfectly collateralized contract 
that has the unit payment of cash in the future time $T$ (See, Eq.~(\ref{perfect_single}).). 
As you can see in ~\cite{collateral_choice} for example, this plays 
the role as the discounting factor for the collateralized cash flow. This fact can be easily 
understood by noticing that there is additivity in price for the perfect and symmetric collateralization~\footnote{There is no point to {\it issue} collateralized zero coupon bond to raise cash from the market since the issuer has to post the same amount of cash as collateral to the buyer.}.

In the same way, for the case where the deal and collateral currencies are different, $(i)$ and $(j)$ respectively,
we {\it define} the foreign collateralized zero coupon bond $D^{(i,j)}$ by
\be
D^{(i,j)}(t,T)=E^{Q^{(i)}}\left[\left. e^{-\int_t^T c^{(i)}_sds}
\left(e^{-\int_t^T y^{(i,j)}_sds}\right)\bold{1}\right|\calf_t\right]~.
\ee
In particular, if $c^{(i)}$ and $y^{(i,j)}$ are independent, we have
\be
D^{(i,j)}(t,T)=D^{(i)}(t,T)e^{-\int_t^T y^{(i,j)}(t,s)ds}~,
\ee
where 
\be
y^{(i,j)}(t,s)=-\frac{\partial}{\partial s}\ln E^{Q^{(i)}}\left[\left.e^{-\int_t^s y_u^{(i,j)}du}\right|\calf_t\right]
\ee
denotes the forward $y^{(i,j)}$ spread.
As before one needs to understand this instrument as the present value of unit payment
collateralized by a foreign currency. 

%%%%%%%%%%%%%%%%%%%%%%%%%%%%%%%%%%%%%%%%%%%%%%%%%%%%%%%%%%%%%%%%%%%%%%%%%%
\subsection{Collateralized FX Forward}
%%%%%%%%%%%%%%%%%%%%%%%%%%%%%%%%%%%%%%%%%%%%%%%%%%%%%%%%%%%%%%%%%%%%%%%%%%
Because of the existence of collateral, FX forward transaction now becomes
non-trivial. The precise understanding of the collateralized FX forward is crucial to 
deal with generic collateralized products.
The definition of currency-$(k)$ collateralized FX forward contract for the 
currency pair $(i,j)$ is as follows~\footnote{In the market, USD is popular as the collateral 
for multi-currency trades.}:
\\
\\
{\it $\bullet$ At the time of trade inception $t$, both parties agree to exchange 
$K$ unit of currency $(i)$ with the one unit of currency $(j)$ at the maturity $T$.
Throughout the contract period, the continuous collateralization by currency $(k)$ is 
performed, i.e. the party who has negative mark-to-market needs to post the 
equivalent amount of cash in currency $(k)$ to the counter party as collateral, 
and this adjustment is made continuously.
The FX forward rate $f_x^{(i,j)}(t,T;k)$ is defined as the value of $K$ that makes
the value of the above contract zero at the time of its inception.}\\\\
By using the results of Sec.~\ref{symmetric}, $K$ needs to satisfy the 
following relation:

\be
K E^{Q^{(i)}}\left[\left. e^{-\int_t^T \left(c_s^{(i)}+y_s^{(i,k)}\right)ds}\right|\calf_t\right]
-f_x^{(i,j)}(t)E^{Q^{(j)}}\left[\left. e^{-\int_t^T\left(c_s^{(j)}+y^{(j,k)}_s\right)ds}\right|\calf_t\right]=0
\ee
and hence the FX forward is given by
\bea
f_x^{(i,j)}(t,T;k)&=&f_x^{(i,j)}(t)\frac{E^{Q^{(j)}}\left[\left. e^{-\int_t^T\left(c_s^{(j)}+y^{(j,k)}_s\right)ds}\right|\calf_t\right]}{E^{Q^{(i)}}\left[\left. e^{-\int_t^T \left(c_s^{(i)}+y_s^{(i,k)}\right)ds}\right|\calf_t\right]}\\
&=&f_x^{(i,j)}(t)\frac{D^{(j,k)}(t,T)}{D^{(i,k)}(t,T)}~,
\eea
which becomes a martingale when $D^{(i,k)}(\cdot, T)$ is used as the numeraire.
In particular, we have
\bea
E^{Q^{(i)}}\left[\left. e^{-\int_t^T \bigl(c_s^{(i)}+y^{(i,k)}_s\bigr)ds}f_x^{(i,j)}(T)
\right|\calf_t\right]&=&D^{(i,k)}(t,T)E^{T^{(i,k)}}\left[\left. f_x^{(i,j)}(T,T;k)\right|\calf_t\right]\nonumber\\
&=&D^{(i,k)}(t,T)f_x^{(i,j)}(t,T;k)~.
\eea
Here, we have defined the $(k)$-collateralized $(i)$ forward measure $T^{(i,k)}$,
where $D^{(i,k)}(\cdot, T)$ is used as the numeraire. $E^{T^{(i,k)}}[\cdot]$ denotes
expectation under this measure.
\\
\\
{\it Remark:} In the case of FX futures, the trade is continuously settled and hence the gain or loss 
is immediately realized. Since the realized gain or loss is not considered as collateral, there is no
exchange of collateral rate $c$ between the financial firm and security exchange. 
Therefore, unless we set $c=0$, the collateralized FX forward 
value is different from that of futures.

%%%%%%%%%%%%%%%%%%%%%%%%%%%%%%%%%%%%%%%%%%%%%%%%%%%%%%%%%%%%%%%%%%%%%%%%%%%
\subsection{Overnight Index Swap}
%%%%%%%%%%%%%%%%%%%%%%%%%%%%%%%%%%%%%%%%%%%%%%%%%%%%%%%%%%%%%%%%%%%%%%%%%%%
The overnight index swap (OIS) is a fixed-vs-floating swap which is the same 
as the usual IRS except that the floating leg pays periodically, say quarterly, 
compounded ON rate instead of Libors.
Let us consider $T_0$-start $T_N$-maturing OIS of currency $(j)$ with fixed rate $S_N$,
where $T_0\geq t$.
If the party $1$ takes a receiver position, we have
\be
dD_s=\sum_{n=1}^N\delta_{T_n}(s)\left[
%\left. 
\Delta_n S_N +1-\exp\left(\int_{T_{n-1}}^{T_n}c^{(j)}_u du\right)
\right]
% \right|\calf_t\right]~,
\ee
where $\Delta$ is day-count fraction of the fixed leg, and $\delta_T(\cdot)$ denotes
Dirac delta function at $T$.
%Using the results of Sec.~\ref{symmetric}, in the case of currency $(k)$ collateralization,
%we have
%\bea
%V_t^{(j)}&=&E^{Q^{(j)}}\left[\left. \int_{]T_0,T_N]}\exp\left(-\int_t^s \bigl(c_u^{(j)}+y_u^{(j,k)}\bigr)du\right)
%dD_s\right|\calf_t\right]\\
%&=&\sum_{n=1}^N E^{Q^{(j)}}\left[\left. e^{-\int_t^{T_n}\bigl(c_u^{(j)}+y_u^{(j,k)}\bigr)du}
%\left(\Delta_n S_N +1-e^{\int_{T_{n-1}}^{T_n} c_u^{(j)}du}  \right)\right|\calf_t \right]~.
%\eea
In particular, if OIS is collateralized by its domestic currency $(j)$, its value $V_t$ is given by
\be
V_t=\sum_{n=1}^N\Delta_n D^{(j)}(t,T_n)S_N-\left(D^{(j)}(t,T_0)-D^{(j)}(t,T_N)\right)~,
\ee
and hence the par rate is expressed as
\be
S_N=\frac{D^{(j)}(t,T_0)-D^{(j)}(t,T_N)}{\sum_{n=1}^N \Delta_n D^{(j)}(t,T_n)}~.
\ee

%%%%%%%%%%%%%%%%%%%%%%%%%%%%%%%%%%%%%%%%%%%%%%%%%%%%%%%%%%%%%%%%%%%%%%%%%%%%%
\subsection{Cross Currency Swap}\label{MtMCCOIS}
%%%%%%%%%%%%%%%%%%%%%%%%%%%%%%%%%%%%%%%%%%%%%%%%%%%%%%%%%%%%%%%%%%%%%%%%%%%%%
Cross currency swap (CCS) is one of the most fundamental products in FX market.
Especially, for maturities longer than a few years, CCS is much more liquid 
than FX forward contract and is the dominant funding source of foreign currencies.
The current market is dominated by USD crosses where 3m USD Libor flat is 
exchanged by 3m Libor of a different currency with additional (negative in many cases) basis spread.
The most popular type of CCS is called MtMCCS (Mark-to-Market CCS) in which 
the notional of USD leg is refreshed at the every Libor fixing time, while the 
notional of the other leg is kept constant throughout the contract.
For model calibration, MtMCCS should be used as we have done in Ref.~\cite{collateral_choice} considering
its liquidity. However, in this paper, we study another type of CCS, which is actually tradable in the market,
to make the link between $y$ and CCS much clearer.

We study the Mark-to-Market cross currency overnight index swap (MtMCCOIS), which is exactly the same as the usual 
MtMCCS except that it pays a compounded ON rate, instead of the Libor, of each currency periodically.
Let us consider $(i,j)$-MtMCCOIS where currency $(i)$ intended to be USD and needs notional refreshments,
and currency $(j)$ is the one in which the basis spread is to be paid.
Let us suppose the party $1$ is the spread receiver and consider $T_0$-start $T_N$-maturing $(i,j)$-MtMCCOIS.
For the $(j)$-leg, we have
\be
dD^{(j)}_s=-\delta_{T_0}(s)+\delta_{T_N}(s)+\sum_{n=1}^N\delta_{T_n}(s)
\left[\left(e^{\int_{T_{n-1}}^{T_n}c^{(j)}_udu}-1\right)+\delta_n B_N\right]~,
\ee 
where $B_N$ is the basis spread of the contract.
For $(i)$-leg, in terms of currency $(i)$, we have
\be
dD^{(i)}_s=\sum_{n=1}^N\left[ \delta_{T_{n-1}}(s)f_x^{(i,j)}(T_{n-1})-
\delta_{T_n}(s)f_x^{(i,j)}(T_{n-1})e^{\int_{T_{n-1}}^{T_n}c^{(i)}_udu}\right]~.
\ee
In total, in terms of currency $(j)$, we have
\bea
dD_s&=&dD_s^{(j)}+f_x^{(j,i)}(s) dD_s^{(i)}\\
&=&dD_s^{(j)}+\sum_{n=1}^N\left[\delta_{T_{n-1}}(s)-\delta_{T_n}(s)\frac{f_x^{(j,i)}(T_n)}{f_x^{(j,i)}(T_{n-1})}
e^{\int_{T_{n-1}}^{T_n}c^{(i)}_udu}\right] \\
&=&\sum_{n=1}^N\delta_{T_n}(s)\left[
e^{\int_{T_{n-1}}^{T_n}c^{(j)}_udu}+\delta_n B_N-\frac{f_x^{(j,i)}(T_n)}{f_x^{(j,i)}(T_{n-1})}e^{\int_{T_{n-1}}^{T_n}
c^{(i)}_udu}\right]~.
\eea

If the collateralization is done by currency $(k)$, then the value for the party $1$
is given by 
\be
V_t=\sum_{n=1}^NE^{Q^{(j)}}\left[\left.
e^{-\int_t^{T_n}(c^{(j)}_u+y_u^{(j,k)})du}\left\{
e^{\int_{T_{n-1}}^{T_n}c^{(j)}_udu}+\delta_n B_N-\frac{f_x^{(j,i)}(T_n)}{f_x^{(j,i)}(T_{n-1})}e^{\int_{T_{n-1}}^{T_n}
c^{(i)}_udu}
\right\}\right|\calf_t\right]~,
\label{CCOIS_k}
\ee
where $T_0\geq t$. 
In particular, if the swap is collateralized by currency $(i)$ (or USD)
that is popular in the market, we obtain 
\bea
V_t&=&\sum_{n=1}^N \delta_n B_N D^{(j)}(t,T_{n})e^{-\int_t^{T_n}y^{(j,i)}(t,u)du}\nonumber\\
&&\quad -\sum_{n=1}^ND^{(j)}(t,T_{n-1})e^{-\int_{t}^{T_{n-1}}y^{(j,i)}(t,u)du}
\left(1-e^{-\int_{T_{n-1}}^{T_n}y^{(j,i)}(t,u)du}\right)\nonumber\\
&=&\sum_{n=1}^N\left[
\delta_n B_N D^{(j,i)}(t,T_n)-D^{(j,i)}(t,T_{n-1})\left(1-e^{-\int_{T_{n-1}}^{T_n}y^{(j,i)}(t,u)du}\right)
\right]~.
\eea
Here, we have assumed the independence of $c^{(j)}$ and $y^{(j,i)}$. 
In fact, the assumption seems reasonable according to the recent 
historical data studied in Ref.~\cite{collateral_choice}.
In this case, we obtain the par MtMCCOIS basis spread as
\be
B_N=\frac{\sum_{n=1}^N D^{(j,i)}(t,T_{n-1})\left(1-
e^{-\int_{T_{n-1}}^{T_n}y^{(j,i)}(t,u)du}\right)}{\sum_{n=1}^N\delta_n D^{(j,i)}(t,T_n)}~.
\ee
Thus, it is easy to see that
\be
B_N\sim \frac{1}{T_N-T_0}\int_{T_0}^{T_N}y^{(j,i)}(t,u)du,
\ee
which gives us the relation between the currency funding spread $y^{(j,i)}$ and 
the observed cross currency basis. Therefore, the cost of collateral $y$ is directly linked to the dynamics of CCS markets.
It is interesting to understand the origin of the funding spread $y^{(i,k)}$ in the pricing formula (\ref{perfect_foreign})
based on the above discussion of CCS. Interested readers can read Appendix~\ref{funding_spread} for the details.

%%%%%%%%%%%%%%%%%%%%%%%%%%%%%%%%%%%%%%%%%%%%%%%%%%%%%%%%%%%%%%%%%%%%%%%%%%%%%%%%%
\section{Numerical Studies for Asymmetric Collateralization}\label{numerical}
%%%%%%%%%%%%%%%%%%%%%%%%%%%%%%%%%%%%%%%%%%%%%%%%%%%%%%%%%%%%%%%%%%%%%%%%%%%%%%%%%
In this section, we study the effects of perfect but asymmetric collateralization and hence 
$(\rm{CVA}=0,~{\rm CCA}\neq 0)$, using the two
fundamental products, MtMCCOIS and OIS. The results will clearly tell us that the 
sophistication of collateral management does matter in the real business.
For both cases, we use the following dynamics in Monte Carlo simulation:
\bea
dc^{(j)}_t&=&\left(\theta^{(j)}(t)-\kappa^{(j)}c^{(j)}_t\right)dt+\sigma_c^{(j)}dW_t^{1}\\
dc^{(i)}_t&=&\left(\theta^{(i)}(t)-\rho_{2,4}\sigma_c^{(i)}\sigma_x^{(j,i)}-\kappa^{(i)}c^{(i)}_t
\right)dt +\sigma_c^{(i)}dW_t^2\\
dy^{(j,i)}_t&=&\left(\theta^{(j,i)}(t)-\kappa^{(j,i)}y^{(j,i)}_t\right)dt+\sigma_y^{(j,i)}dW_t^3\\
d\ln f_{x}^{(j,i)}(t)&=&\left(c^{(j)}_t-c^{(i)}_t+y^{(j,i)}_t-\frac{1}{2}(\sigma_x^{(j,i)})^2\right)dt
+\sigma_x^{(j,i)}dW_t^4
\eea
where $\{W^i, i=1\cdots 4\}$ are Brownian motions under the spot martingale measure of currency $(j)$.
Every $\theta(t)$ is a deterministic function of time, and is adjusted in such a way that 
we can recover the initial term structures of the relevant variable. The 
procedures for the curve construction are given in Appendix~\ref{curve_construction}.
We assume every $\kappa$ and $\sigma$ are constants.
We allow general correlation structure $(d[W^i,W^j]_t=\rho_{i,j}dt)$ except that $\rho_{3,j}=0$ for all $j\neq 3$.

The above dynamics is chosen just for simplicity and demonstrative purpose, 
and generic HJM framework can also be applied to 
the evaluation of Gateaux derivative. For details of generic dynamics in 
HJM framework, see Refs.~\cite{dynamic_basis, collateral_termstructures}. In the following, we use the term structure 
for the $(i,j)$ pair taken from the typical data of $(\rm{USD,JPY})$ in early 2010 for presentation. 
In Appendix~\ref{Data}, we have provided the term structures and other parameters used in simulation.

%The discussed form of asymmetry is particularly interesting, since even if the 
%relevant CSA is actually symmetric, the asymmetry arises effectively if there is difference 
%in the level of sophistication of collateral management. From the following two examples,
%one can see that the efficient collateral management is practically relevant and 
%the firms who are incapable of doing so will have to pay quite expensive cost to the 
%counter party, and vice versa.
%%%%%%%%%%%%%%%%%%%%%%%%%%%%%%%%%%%%%%%%%%%%%%%%%%%%%%%%%%%%%%%%%
\subsection{ Asymmetric Collateralization for MtMCCOIS }
%%%%%%%%%%%%%%%%%%%%%%%%%%%%%%%%%%%%%%%%%%%%%%%%%%%%%%%%%%%%%%%%%
We now implement Gateaux derivative using Monte Carlo simulation based on the 
model we have just explained.
To see the accuracy of Gateaux derivative, we have compared it with 
a numerical result directly obtained from PDE using a simplified setup in Appendix~\ref{PDE}.
\begin{figure}
	\center{\includegraphics[width=125mm]{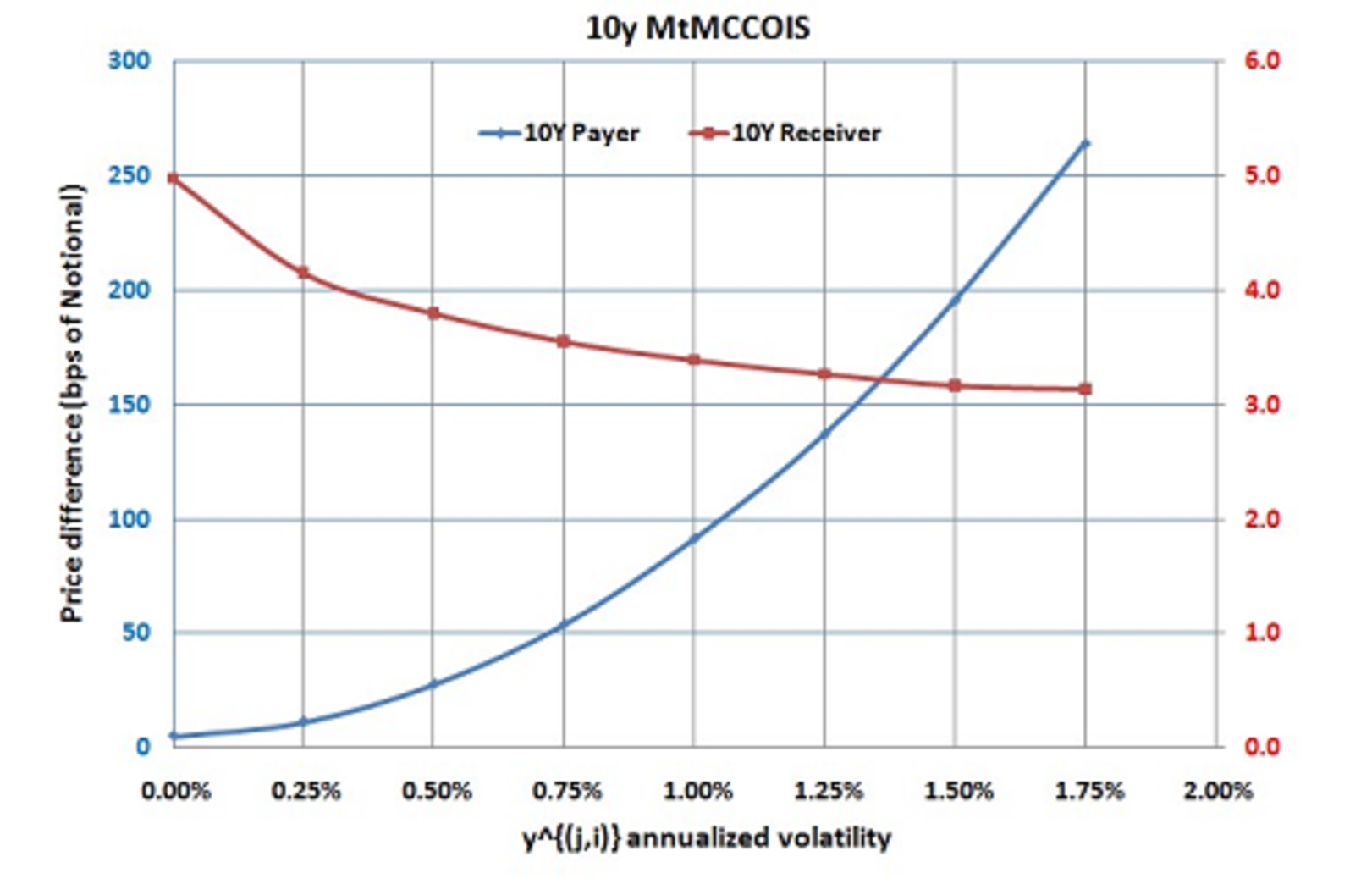}}
	\vspace{0mm}
	\caption{Price difference from symmetric limit for 10y MtMCCOIS}
	\label{10yCCS}
	\center{\includegraphics[width=120mm]{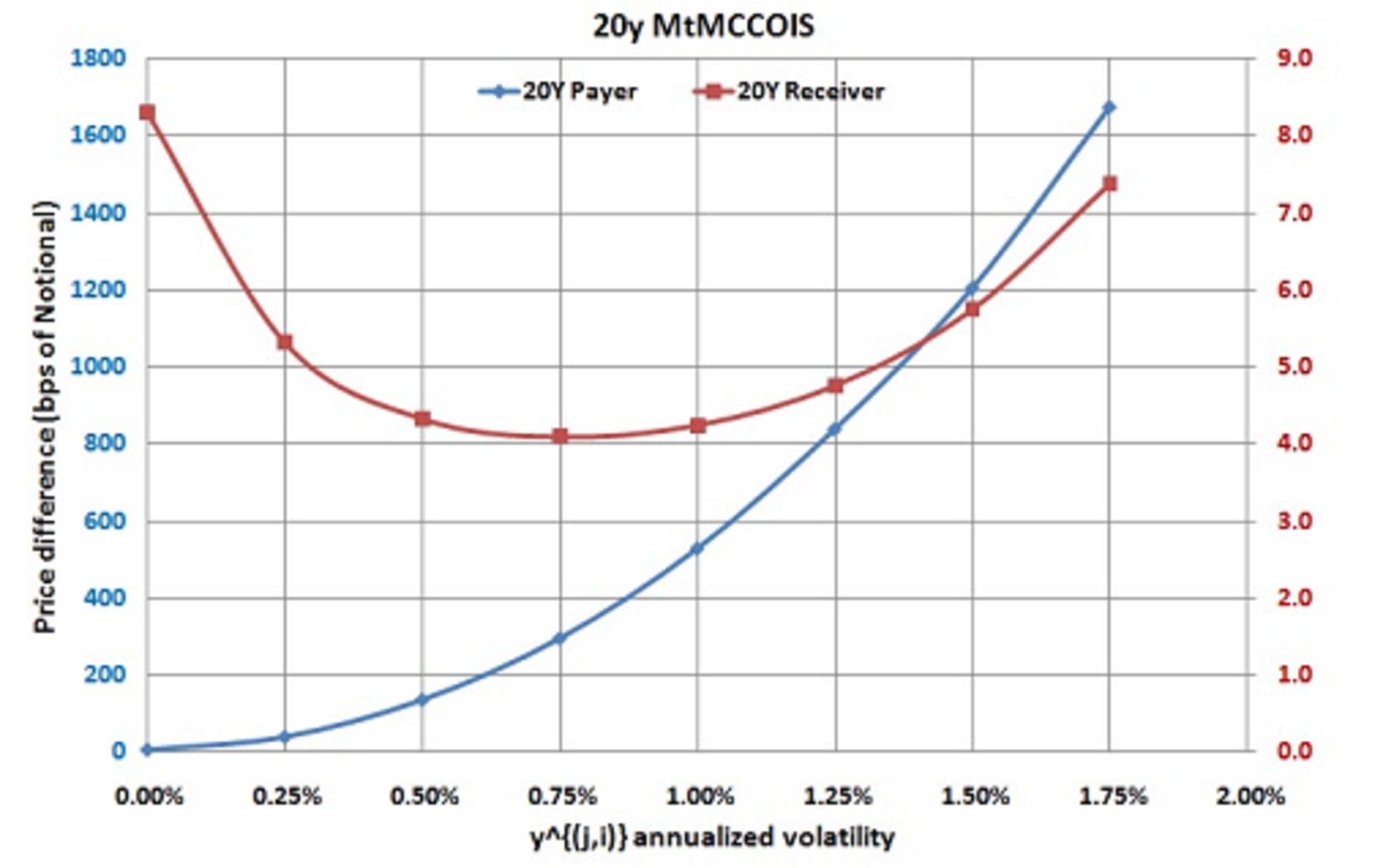}}
	\vspace{0mm}
	\caption{Price difference from symmetric limit for 20y MtMCCOIS}
	\label{20yCCS}
\end{figure}

Firstly, we consider  MtMCCOIS explained in Sec.~\ref{MtMCCOIS}.
We consider a spot-start, $T_N$-maturing $(i,j)$-MtMCCOIS, where the leg of currency $(i)$ (intended to be USD) needs notional 
refreshments. 
Let us assume perfect but asymmetric collateralization as follows: \\
(1) Party $1$  can use either the currency $(i)$ or $(j)$ as collateral.\\
(2) Party $2$  can only use the currency $(i)$ as collateral. \\
For the derivation of the present value, see Appendix~\ref{asymmetric_MtMCCOIS}.

In Figs.~\ref{10yCCS} and \ref{20yCCS}, we have shown the numerical result of {\rm CCA}
which is the price difference from the symmetric limit, for 10y and 20y MtMCCOIS, respectively.
The spread $B$ was chosen in such way that the value in symmetric limit, $\overline{V}_0$, becomes zero.
In both cases, the horizontal axis is the annualized volatility of $y^{(j,i)}$, and 
the vertical one is the price difference from ${\rm CCA}$ in terms of bps of notional of currency $(j)$.
When the party $1$ is the spread payer (receiver), we have used the left (right) axis.
The results are rather insensitive to the FX volatility due to the notional refreshments 
of currency-$(i)$ leg.  From the historical analysis performed in Ref.~\cite{collateral_choice},
we know that annualized volatility of $y^{(j,i)}$ tends to be $50$bps or so in a calm market, but 
it can be $(100\sim 200)$bps or more in a volatile market for major currency pairs, such as (EUR,USD) and (USD, JPY).
Therefore, the impact of asymmetric collateralization in this example
can be practically very significant when party $1$ is the spread payer.
When the party $1$ is the spread receiver, one sees that the impact of asymmetry is very small, only a few bps of notional.
This can be easily understood in the following way: When the party $1$ has a negative mark-to-market and hence 
is the collateral payer who has the option to 
change the collateral currency, $y^{(j,i)}$ tends to be large and hence the optimal currency remains the 
same currency $(i)$.  

Finally, let us briefly mention about the standard MtMCCS with Libor payments.
As discussed in Ref.~\cite{collateral_choice}, the contribution from Libor-OIS spread to CCS is not significant 
relative to that of $y^{(j,i)}$. Therefore, the numerical significance of asymmetric collateralization 
is expected to be quite similar in the standard case, too. 

%\clearpage
%%%%%%%%%%%%%%%%%%%%%%%%%%%%%%%%%%%%%%%%%%%%%%%%%%%%%%%%%%%%%%%%%%%%%%%%%%%%
\subsection{Asymmetric Collateralization for OIS}
%%%%%%%%%%%%%%%%%%%%%%%%%%%%%%%%%%%%%%%%%%%%%%%%%%%%%%%%%%%%%%%%%%%%%%%%%%%%
Now we study the impact of asymmetric collateralization on OIS.
We consider OIS of currency $(j)$, and assume the following asymmetry in collateralization:\\
(1) Party $1$  can use either the currency $(i)$ or $(j)$ as collateral.\\
(2) Party $2$  can only use the currency $(j)$ (domestic currency) as collateral.\\
For the derivation of the present value, see Appendix~\ref{asymmetric_OIS}.

\begin{figure}
\vspace{-14mm}
	\center{\includegraphics[width=112mm]{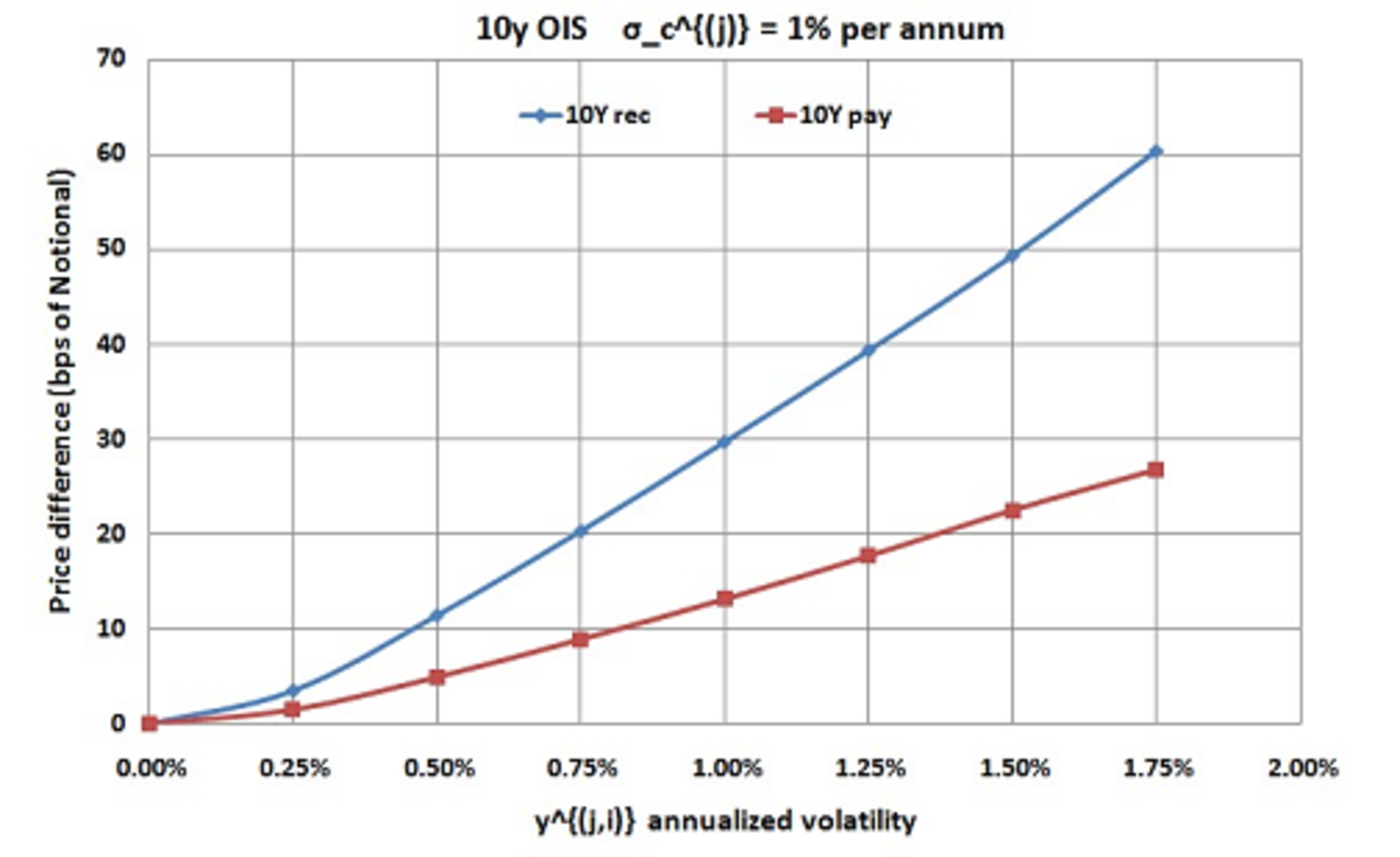}}
	\vspace{-5mm}
	\caption{Price difference from symmetric limit for 10y OIS}
	\label{10yOIS}

	\center{\includegraphics[width=112mm]{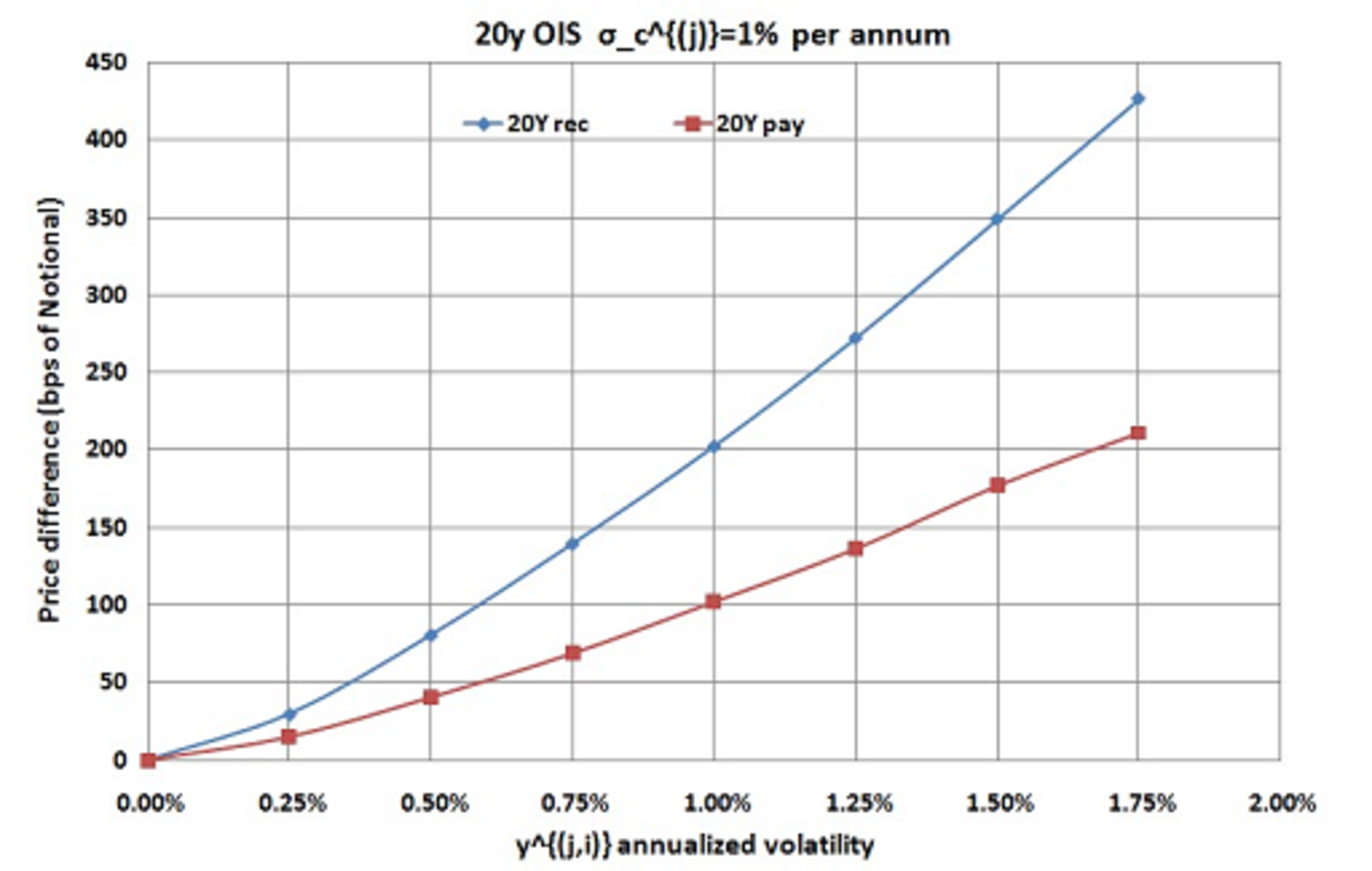}}
	\vspace{-5mm}
	\caption{Price difference from symmetric limit for 20y OIS}
	\label{20yOIS}

	\center{\includegraphics[width=112mm]{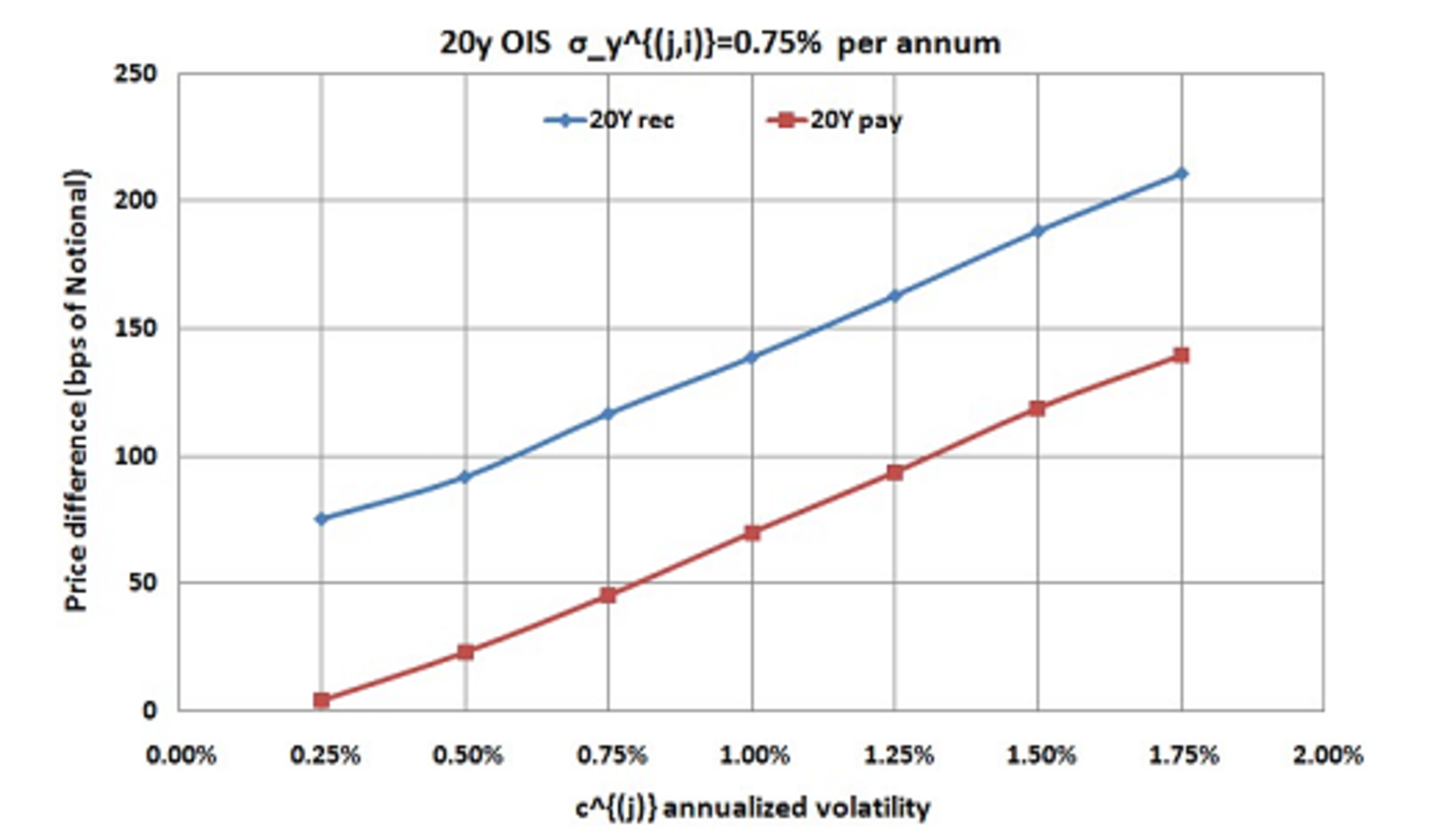}}
	\vspace{-5mm}
	\caption{Price difference from symmetric limit for 20y OIS for the change of $\sigma_c^{(j)}$ }
	\label{20yCVOL}
\end{figure}

In Figs.~\ref{10yOIS}, \ref{20yOIS}, and \ref{20yCVOL}, we have shown the numerical results of ${\rm CCA}$
for 10y and 20y OIS from the view point of party $1$. In the first two figures, we have fixed $\sigma_c^{(j)}=1\%$ and changed 
$\sigma_y^{(j,i)}$ to see the sensitivity against CCS. In the last figure, we have fixed 
the $y^{(j,i)}$ volatility as $\sigma_y^{(j,i)}=0.75\%$ and changed the volatility of collateral rate $c^{(j)}$.
Since the term structure of OIS used in simulation is upward sloping, 
the mark-to-market value of the fixed receiver tends to be negative 
in the long end of the contract. This makes the cheapest-to-deliver optionality bigger for the 
receiver, and hence it has bigger {\rm CCA} contribution than the case of payer.

%%%%%%%%%%%%%%%%%%%%%%%%%%%%%%%%%%%%%%%%%%%%%%%%%%%%%%%%%%%%%%%%%
\section{General Implications of Asymmetric Collateralization}
\label{asymmetric_implications}
%%%%%%%%%%%%%%%%%%%%%%%%%%%%%%%%%%%%%%%%%%%%%%%%%%%%%%%%%%%%%%%%%
From the results of section~\ref{numerical}, we have seen the practical significance of 
asymmetric collateralization. It is now clear that sophisticated financial firms 
may obtain significant funding benefit from the less-sophisticated counter parties.

Before going to discuss the imperfect collateralization and associated CVA, 
let us explain two generic implications of asymmetric collateralization,
one for netting and the other for resolution of information, which is closely 
related to the observation just explained.
Although derivation itself can be done in exactly the same way as Ref.~\cite{Duffie}
after the reinterpretation of several variables, we get new insights for collateralization
that may be important for the appropriate design and regulations for 
the financial market.

%%%%%%%%%%%%%%%%%%%%%%%%%%%%%%%%%%%%%%%%%%%%%%%%%%%%%%%%%%%%%%%%%%%%
\subsection{An implication for Netting}
%%%%%%%%%%%%%%%%%%%%%%%%%%%%%%%%%%%%%%%%%%%%%%%%%%%%%%%%%%%%%%%%%%%%
\begin{proposition}\footnote{We assume perfect collateralization just for clearer interpretation.
The results will not change qualitatively as long as 
$\delta^iy_t^i > (1-R_t^i)(1-\delta_t^i)^+h_t^i-(1-R_t^j)(\delta_t^i-1)^+h_t^j$.}
Assume perfect collateralization. Suppose that, for each party $i$, $y_t^i$ is bounded 
and does not depend on the contract value directly. Let $V^a$, $V^b$, and $V^{ab}$ be, respectively, 
the value processes (from the view point of party $1$) of contracts with cumulative dividend processes
$D^a$, $D^b$, and $D^a+D^b$ (i.e., netted portfolio).  If $y^1\geq y^2$, then $V^{ab}\geq V^a+V^b$, and if 
$y^1\leq y^2$, then $V^{ab}\leq V^a+V^b$.
\label{prop-2}
\end{proposition}

Proof is available in Appendix~\ref{proof-2}.
The interpretation of the proposition is very clear: The party who has the higher funding cost $y$ due to 
asymmetric CSA or lack of sophistication in collateral management prefer to have netting agreements 
to decrease funding cost. On the other hand, an advanced financial firm who has capability 
to carry out optimal collateral strategy to achieve the lowest possible value of $y$ tries to 
avoid netting to exploit funding benefit. For example, an advanced firm may prefer to enter an opposite trade with a 
different counterparty rather than to unwind the original trade. For standardized products traded through 
CCPs, such a firm may prefer to use several clearing houses cleverly to avoid netting.

The above finding seems slightly worrisome for the healthy development of CCPs.
Advanced financial firms that have sophisticated financial technology and operational system
are usually primary members of CCPs, and some of them are trying to set up their own 
clearing service facility. If those firms try to exploit funding benefit, they 
avoid concentration of their contracts to major CCPs and may create very disperse
interconnected trade networks and may reduce overall netting opportunity in the market.
Although remaining credit exposure is very small as long as
collateral is successfully being managed, the dispersed use of CCPs may worsen the systemic risk
once it fails. In the work of Duffie \& Huang~\cite{Duffie}, the corresponding 
proposition is derived in the context of bilateral CVA. We emphasize that one important practical difference is the 
strength of incentives provided to financial firms. Although it is somewhat obscure
how to realize profit/loss reflected in CVA, it is rather straightforward in the case of 
collateralization by making use of CCS market as we have explained in Sec.~\ref{MtMCCOIS}.

%%%%%%%%%%%%%%%%%%%%%%%%%%%%%%%%%%%%%%%%%%%%%%%%%%%%%%%%%%%%%%%%%%%%%%%%%%
\subsection{An implication for Resolution of Information}
%%%%%%%%%%%%%%%%%%%%%%%%%%%%%%%%%%%%%%%%%%%%%%%%%%%%%%%%%%%%%%%%%%%%%%%%%%
We once again follow the setup given in Ref~\cite{Duffie}.
We assume the existence of two markets: One is market $F$, which has filtration $\mathbb{F}$,
that is the one we have been studying. The other one is market $G$ with filtration 
$\mathbb{G}=\{\calg_t : t\in[0,T]\}$. The market $G$ is identical to the market $F$ except that 
it has earlier resolution of uncertainty, or in other words, $\calf_t\subseteq \calg_t$ for all $t\in[0,T]$ while
$\calf_0=\calg_0$. The spot marting measure $Q$ is assumed to apply to the both markets.

\begin{proposition}\footnote{We assume perfect collateralization just for clearer interpretation.
The results will not change qualitatively as long as 
$\delta^iy_t^i > (1-R_t^i)(1-\delta_t^i)^+h_t^i-(1-R_t^j)(\delta_t^i-1)^+h_t^j$.}
Assume perfect collateralization. Suppose that, for each party $i$, $y^i$ is bounded and does not 
depend on the contract value directly. Suppose that $r$, $y^1$ and $y^2$ are adapted to both the 
filtrations $\mathbb{F}$ and $\mathbb{G}$. The contract has cumulative 
dividend process $D$, which is a semimartingale of integrable variation with respect to filtrations
$\mathbb{F}$ and $\mathbb{G}$. Let $V^{F}$ and $V^{G}$ denote, respectively, the values of the contract
in markets $F$ and $G$ from the view point of party $1$. 
If $y^1\geq y^2$, then $V_0^{F}\geq V_0^{G}$, and if $y^1\leq y^2$, then $V_0^F\leq V_0^G$.
\label{prop-3}
\end{proposition}

Proof is available in Appendix~\ref{proof-3}.
The proposition implies that the party who has the higher effective funding cost $y$ either from 
the lack of sophisticated collateral management technique or from asymmetric CSA would like to delay 
the information resolution to avoid timely margin call from the counterparty. 
The opposite is true for advanced financial firms which are likely to have
advantageous CSA and/or sophisticated system. The incentives to obtain funding benefit will
urge these firms to provide mark-to-market information of contracts to counter parties in 
timely manner, and seek early resolution of valuation dispute to achieve 
funding benefit. Considering the privileged status of these firms, the latter effects will 
probably be dominant in the market.

%%%%%%%%%%%%%%%%%%%%%%%%%%%%%%%%%%%%%%%%%%%%%%%%%%%%%%%%%%%%%%%%%
\section{Imperfect Collateralization and CVA}
\label{imp_collateral}
%%%%%%%%%%%%%%%%%%%%%%%%%%%%%%%%%%%%%%%%%%%%%%%%%%%%%%%%%%%%%%%%%
As we have explained in Sec.~\ref{sec_generic}, our framework 
can also handle the imperfectly collateralized contract, where there remain 
counter party credit risk as well as collateral cost adjustment.
In the remainder of this paper, we study several important examples of 
imperfect collateralization by using the generic results of Eqs. (\ref{generic_cca}) and (\ref{generic_cva}).
\\
\\
{$\rm{\bold{Case~1}}$}:
Consider the situation where the both parties use collateral currency $(i)$, which is 
the same as the deal currency. 
In this case, CCA and CVA are given by
\bea
{\rm CCA}_t=E^{Q^{(i)}}\left[\left. \int_t^T e^{-\int_t^s c^{(i)}_u du}y^{(i)}_s
\Bigl( -(\delta_s^1 -1)\bigl[-\overline{V}_s\bigr]^+
+(\delta_s^2 -1)\bigl[\overline{V}_s\bigr]^+\Bigr)ds\right|\calf_t\right]
\eea
and 
\bea
{\rm CVA}_t&=&E^Q\left[\left. \int_t^T e^{-\int_t^s c^{(i)}_udu}
h_s^1(1-R_s^1)\Bigl\{(1-\delta_s^1)^+\bigl[-\overline{V}_s]^+
+(\delta_s^2-1)^+\bigl[\overline{V}_s\bigr]^+\Bigr\}ds\right|\calf_t\right] \nn \\
&-&E^Q\left[\left. \int_t^T e^{-\int_t^s c^{(i)}_udu}h_s^2(1-R_s^2)
\Bigl\{(1-\delta_s^2)^+\bigl[\overline{V}_s\bigr]^++
(\delta_s^1-1)^+\bigl[-\overline{V}_s\bigr]^+\Bigr\}ds\right|\calf_t\right]~\nn\\
\eea
where 
\be
\overline{V}_t=E^{Q^{(i)}}\left[\left.\int_{]t,T]}\exp\left(-\int_t^s c^{(i)}_udu\right)dD_s\right|\calf_t\right]
\label{bench_1}
\ee
is a value under perfect collateralization by domestic currency. This would be the most common situation 
in the market. Note that the discounting rate is given by the collateral rate and there also exists CCA
term.
\\
\\
{$\rm{\bold{Case~2}}$}:
Consider the situation where the both parties have to use collateral currency $(j)$
which is different from the deal currency $(i)$.
In this case, CCA and CVA are given by
\bea
{\rm CCA}_t=E^{Q^{(i)}}\left[\left. \int_t^T e^{-\int_t^s \bigl(c^{(i)}_u+y_u^{(i,j)}\bigr) du}y^{(j)}_s
\Bigl( -(\delta_s^1 -1)\bigl[-\overline{V}_s\bigr]^+
+(\delta_s^2 -1)\bigl[\overline{V}_s\bigr]^+\Bigr)ds\right|\calf_t\right]
\eea
and 
\bea
{\rm CVA}_t&=&E^{Q^{(i)}}\left[\left. \int_t^T e^{-\int_t^s \bigl(c^{(i)}_u+y^{(i,j)}_u\bigr)du}
h_s^1(1-R_s^1)\Bigl\{(1-\delta_s^1)^+\bigl[-\overline{V}_s]^+
+(\delta_s^2-1)^+\bigl[\overline{V}_s\bigr]^+\Bigr\}ds\right|\calf_t\right] \nn \\
&-&E^{Q^{(i)}}\left[\left. \int_t^T e^{-\int_t^s \bigl(c^{(i)}_u+y^{(i,j)}_u\bigr)du}h_s^2(1-R_s^2)
\Bigl\{(1-\delta_s^2)^+\bigl[\overline{V}_s\bigr]^++
(\delta_s^1-1)^+\bigl[-\overline{V}_s\bigr]^+\Bigr\}ds\right|\calf_t\right]~\nn\\
\eea
where 
\be
\overline{V}_t=E^{Q^{(i)}}\left[\left.\int_{]t,T]}\exp\left(-\int_t^s \bigl(c^{(i)}_u+y^{(i,j)}_u\bigr)du\right)dD_s\right|\calf_t\right]
\ee
is a value under perfect collateralization by the foreign currency. 
This is also a common situation 
for multi- or non-G5 currency trades collateralized by USD but with sizable uncollateralized exposure due to price disputes,
for example.

Note that the correlation between the currency funding spread $y^{(i,j)}$ 
and hazard rates may contribute significantly to the value of CVA.
This is easy to understand, for example,  by considering the USD collateralized EUR derivatives with an European bank 
as a counter party. As clearly seen in the ongoing turmoil of Euro zone, expensive funding cost of USD 
reflected by widening EUR/USD CCS basis spread seems highly correlated to the 
deteriorating creditworthiness of European banks.
\\
\\
{$\rm{\bold{Case~3}}$}:
Let us consider another important situation, which is the unilateral collateralization with bilateral default risk.
Suppose the setup in $\rm{\bold{Case~1}}$ except that only the party $1$ has to post the collateral due to its 
low creditworthiness relative to the party $2$. In this case, we have
\bea
{\rm CCA}_t=-E^{Q^{(i)}}\left[\left. \int_t^T e^{-\int_t^s c^{(i)}_u du}y^{(i)}_s
\Bigl((\delta_s^1-1)\bigl[-\overline{V}_s\bigr]^+
+\bigl[\overline{V}_s\bigr]^+\Bigr)ds\right|\calf_t\right]
\eea
and 
\bea
{\rm CVA}_t&=&E^{Q^{(i)}}\left[\left. \int_t^T e^{-\int_t^s c^{(i)}_udu}
h_s^1(1-R_s^1)\Bigl\{(1-\delta_s^1)^+\bigl[-\overline{V}_s]^+\Bigr\}ds\right|\calf_t\right] \nn \\
&-&E^{Q^{(i)}}\left[\left. \int_t^T e^{-\int_t^s c^{(i)}_udu}h_s^2(1-R_s^2)
\Bigl\{\bigl[\overline{V}_s\bigr]^++(\delta_s^1-1)^+\bigl[-\overline{V}_s\bigr]^+\Bigr\}ds\right|\calf_t\right]~\nn\\
\eea
where $\overline{V}$ is the same as Eq.~(\ref{bench_1}).
If party $1$ is required to post "strong" currency (that is the currency with high value of $y^{(i)}$, such as USD~\cite{collateral_choice}), and also imposed stringent collateral management $\delta^1\simeq 1$, 
it may suffer significant funding cost from CCA even when CVA is small enough due to the high credit worthiness of the 
party $2$.

Note that, this example is particularly common when SSA (sovereign, supranational and agency) is involved
as party $2$. For example, when it is a central bank, it does not post collateral but receives it.
For the party $1$, it is very difficult to hedge this position.
Typically, the risk associated with the funding cost in {\rm CCA} remains un-hedged,
since party $1$ has to follow bilateral collateralization 
when it enters an offsetting trade in the interbank market.
Once the market interest rate starts to go up while the overnight rate $c$ is kept low by the central bank,
the resultant mark-to-market loss from {\rm CCA} can be quite significant due to the 
rising cost of collateral "$y$".
\\
\\
$\rm{\bold{Case~4}}$:
Our framework can also handle the trades with non-zero collateral thresholds, where margin 
call occurs only when the exposure to the counter party exceeds the threshold. 
A threshold is a level of exposure below which collateral will not be called, and hence 
it represents the amount of uncollateralized exposure. If the exposure is above the threshold,
only the incremental exposure will be collateralized.

Usually, the collateral thresholds are set according to the credit standing of each counter party.
They are often asymmetric, with lower-rated counter party having a lower threshold than the other. 
It may be adjusted during the contract according to the "triggers" linked to the credit ratings
of the contracting parties. We assume that the threshold of counter party $i$ at time $t$ is set by $\Gamma_t^i>0$, and that the 
exceeding exposure is perfectly collateralized continuously.

In the presence of thresholds, Eq.~(\ref{generic_pricing1}) is modified in the following way:
\bea
S_t&=&\beta_tE^Q\left[\left. \int_{]t,T]}\beta_u^{-1}\bold{1}_{\{\tau>u\}}\left\{
dD_u+q(u,S_u)S_u du\right\}\right.\right. \nonumber\\
&&\quad +\left.\left.\int_{]t,T]}\beta_u^{-1}\bold{1}_{\{\tau\geq u\}}
\left\{ Z^1(u,S_{u-})dH_u^1+Z^2(u,S_{u-})dH_u^2\right\}
\right|\calf_t\right]~,
\eea
where
\bea
q(t,S_t)=y_t^1\left(1+\frac{\Gamma_t^1}{S_t}\right)\bold{1}_{\{S_t<-\Gamma_t^1\}}
+y_t^2\left(1-\frac{\Gamma_t^2}{S_t}\right)\bold{1}_{\{S_t\geq \Gamma_t^2\}}~,
\eea
and
\bea
Z^1(t,S_t)&=&S_t\left[ \left(1+(1-R_t^1)\frac{\Gamma_t^1}{S_t}\right)\bold{1}_{\{S_t<-\Gamma^1_t\}}
+R^1_t\bold{1}_{\{-\Gamma_t^1\leq S_t< 0\}}+\bold{1}_{\{S_t\geq 0\}}
\right]~\nonumber\\
Z^2(t,S_t)&=&S_t\left[\left(1-(1-R_t^2)\frac{\Gamma_t^2}{S_t}\right)\bold{1}_{\{S_t\geq \Gamma_t^2\}}
+R_t^2\bold{1}_{\{0\leq S_t< \Gamma_t^2\}}+\bold{1}_{\{S_t< 0\}}
\right] \nonumber~.
\eea
Here, we have assumed the same recovery rate for the uncollateralized exposure 
regardless of whether the contract value is above or below the threshold.

Following the same procedures given in Proposition~\ref{prop-1}, one can show that the 
pre-default value of the contract $V_t$ is given by
\be
V_t=E^Q\left[\left.\int_{]t,T]}\exp\left(-\int_t^s \bigl(r_u-\mu(u,V_u)\bigr)du\right)dD_s\right|\calf_t\right]~,\quad t\leq T
\ee
where
\bea
\mu(t,V_t)&=&y_t^1\bold{1}_{\{V_t<0\}}+y_t^2\bold{1}_{\{V_t\geq 0\}}\nonumber\\
&&-\left(y_t^1+h_t^1(1-R_t^1)\right)
\left[\bold{1}_{\{-\Gamma_t^1\leq V_t<0\}}-\frac{\Gamma_t^1}{V_t}\bold{1}_{\{V_t<-\Gamma_t^1\}}\right]\nonumber\\
&&-\left(y_t^2+h_t^2(1-R_t^2)\right)\left[\bold{1}_{\{0\leq V_t < \Gamma_t^2\}}+\frac{\Gamma_t^2}{V_t}\bold{1}_{\{V_t\geq \Gamma_t^2\}}
\right]~.
\eea

Now, consider the case where the both parties use the same collateral currency $(i)$, which is equal to the
deal currency. Then, we have
\bea
\mu(t,V_t)&=&y_t^{(i)}-\l\{ y_t^{(i)}\bold{1}_{\{-\Gamma_t^1\leq V_t< \Gamma_t^2\}}\r.\nonumber\\
&&+y_t^{(i)}\left[\frac{\Gamma_t^1}{V_t}\bold{1}_{\{V_t<-\Gamma_t^1\}}-\frac{\Gamma_t^2}{V_t}\bold{1}_{\{V_t\geq \Gamma_t^2\}}
\right]
\nonumber\\
&&-h_t^1(1-R_t^1)\left[\bold{1}_{\{-\Gamma_t^1\leq V_t<0\}}-\frac{\Gamma_t^1}{V_t}\bold{1}_{\{V_t<-\Gamma_t^1\}}\right]
\nonumber\\
&&
\l.
-h_t^2(1-R_t^2)\left[\bold{1}_{\{0\leq V_t<\Gamma_t^2\}}+\frac{\Gamma_t^2}{V_t}\bold{1}_{\{V_t\geq \Gamma_t^2\}}
\right]
\r\}.
\eea
Thus, applying Gateaux derivative around the symmetric perfect collateralization with currency $(i)$ i.e. 
$\overline{y}=y^{(i)}$, we obtain
\be
V_t\simeq \overline{V}_t+{\rm CCA}+{\rm CVA},
\ee
where
\be
\overline{V}_t=\E^{Q^{(i)}}\left[\left.\int_{]t,T]}\exp\left(-\int_t^s c^{(i)}_u du\right)dD_s\right|\calf_t\right]~,
\ee
and
\bea
{\rm CCA}_t&=&-E^{Q^{(i)}}\left[\left.\int_t^T e^{-\int_t^s c_u^{(i)}du}y_s^{(i)}\overline{V}_s
\bold{1}_{\{-\Gamma_s^1\leq \overline{V}_s
< \Gamma_s^2\}}ds\right|\calf_t\right]\nonumber\\
&&+E^{Q^{(i)}}\left[\left.\int_t^T e^{-\int_t^s c_u^{(i)}du}
y_s^{(i)}\bigl[\Gamma_s^1\bold{1}_{\{\overline{V}_s<-\Gamma_s^1\}}-\Gamma_s^2
\bold{1}_{\{\overline{V}_s\geq \Gamma_s^2\}}\bigr]ds
\right|\calf_t\right]\\
{\rm CVA}_t&=&-E^{Q^{(i)}}\left[\left.\int_t^Te^{-\int_t^s c_u^{(i)}du}\Bigl[
h_s^1(1-R_s^1)\bigl(\overline{V}_s\bold{1}_{\{-\Gamma_s^1\leq \overline{V}_s<0\}}-\Gamma_s^1
\bold{1}_{\{\overline{V}_s<-\Gamma_s^1\}}
\bigr)\Bigr]ds\right|\calf_t\right]\nonumber\\
&&-E^{Q^{(i)}}\left[\left.\int_t^Te^{-\int_t^s c_u^{(i)}du}
\Bigl[h_s^2(1-R_s^2)\bigl(\overline{V}_s\bold{1}_{\{0<\overline{V}_s\leq \Gamma_s^2\}}
+\Gamma_s^2\bold{1}_{\{\overline{V}_s>\Gamma_s^2\}}\bigr)\Bigr]ds\right|\calf_t\right]~.\nonumber\\
\eea
It is easy to see that the terms in CCA are reflecting the fact that no collateral is being posted in 
the range $\{-\Gamma_t^1\leq V_t\leq \Gamma_t^2\}$, and that 
the posted amount of collateral is smaller than $|V|$ by the size of threshold.
The terms in CVA represent bilateral uncollateralized credit exposure, which is capped by each threshold.
\\
\\
%%%%%%%%%%%%%%%%%%%%%%%%%%%%%%%%%%%%%%%%%%%%%%%%%%%%%%
{\it Remarks:}
%%%%%%%%%%%%%%%%%%%%%%%%%%%%%%%%%%%%%%%%%%%%%%%%%%%%%%
Although we have treated $(\delta^i_t)$ as the process of the  collateral coverage ratio, it is also useful to 
handle the collateral devaluation. It is plausible that the value of collateral is highly linked 
to the counter party and its value may jump downward at the time of default.
For example, we can consider a USD interest rate swap collateralized by EUR cash with an European
bank as a counter party. It is easy to imagine that EUR/USD jump downward at the time of default
of the European bank. If we assume the bilateral perfect collateralization, then ${\rm CCA}$ is zero,
but there appears non-zero $\rm{CVA}$ which can be calculated by interpreting $\delta_{\tau}^i$ as the fraction of 
devaluation of the collateral posted by party $i$. We can introduce the collateral coverage ratio and
the fraction of collateral devaluation separately to handle more generic situations.

%%%%%%%%%%%%%%%%%%%%%%%%%%%%%%%%%%%%%%%%%%%%%%%%%%%%%%%%%%%%%%%%%%%%%%%%%%%
\section{Conclusions}
\label{conclusion}
%%%%%%%%%%%%%%%%%%%%%%%%%%%%%%%%%%%%%%%%%%%%%%%%%%%%%%%%%%%%%%%%%%%%%%%%%%%
This article develops the methodology to deal with asymmetric and 
imperfect collateralization as well as remaining counterparty credit risk.
It was shown that all of the issues are able to be handled in an unified way 
by making use of Gateaux derivative. We have shown that the 
resulting formula contains CCA that represents adjustment of collateral cost 
due to the deviation from the perfect collateralization, and the terms corresponding to the bilateral CVA.
The credit value adjustment now contains the possible dependency among cost of collaterals, hazard rates,
collateral coverage ratio and the underlying contract value.
Even if we assume that the collateral coverage ratio and recovery rate are constant, 
the change of effective discounting rate induced by the currency funding spread and its correlation 
to the hazard rates may significantly change the size of the adjustment.

Direct link of CCS spread and collateral cost allows us to study the 
numerical significance of asymmetric collateralization.
From the numerical analysis using CCS and OIS, 
the relevance of sophisticated collateral management is now clear.
If a financial firm is incapable of posting the cheapest collateral currency, 
it has to pay very expensive funding cost to the counter party.
We also explained the issue of one-way CSA, which is common when SSA entities are involved.
If the funding cost of collateral (or "$y$") rises,
the financial firm that is the counterparty of SSA may suffer from significant 
mark-to-market loss from CCA, and it is quite difficult to hedge.

The article also discussed some generic implications of collateralization.
In particular, it was shown that the sophisticated financial firms have an incentive to avoid 
netting of trades if they try to exploit funding benefit as much as possible, which may
reduce the overall netting opportunity and potentially increase
the systemic risk in the financial market.

%%%%%%%%%%%%%%%%%%%%%%%%%%%%%%%%%%%%%%%%%%%%%%%%%%%%%%%%%%%%%%%%%%%%%%%%%%%%
\appendix
%%%%%%%%%%%%%%%%%%%%%%%%%%%%%%%%%%%%%%%%%%%%%%%%%%%%%%%%%%%%%%%%%%%%%%%%%%%%
\section{ Proof of Proposition~\ref{prop-1}}\label{proof-1}
%%%%%%%%%%%%%%%%%%%%%%%%%%%%%%%%%%%%%%%%%%%%%%%%%%%%%%%%%%%%%%%%%%%%%%%%%%%%
Firstly, we consider the SDE for $S_t$. Let us define $L_t=1-H_t$. One can show that 
\bea
&&\beta_t^{-1}S_t+\int_{]0,t]}\beta_u^{-1}L_u\bigl(dD_u+q(u,S_u)S_u du\bigr)
+\int_{]0,t]}\beta_u^{-1}L_{u-}
\bigl(Z^1(u,S_{u-})dH_u^1+Z^2(u,S_{u-})dH_u^2\bigr)\nonumber\\
&&=\E^Q\left[\int_{]0,T]}\beta_u^{-1}\bold{1}_{\{\tau>u\}}
\Bigl\{ dD_u + \left(y_u^{1}\delta_u^1\bold{1}_{\{S_{u}<0\}}+
y_u^2 \delta_u^2 \bold{1}_{\{S_{u}\geq 0\}}\right)S_{u}du\Bigr\}\right. \nonumber\\
&&\qquad\qquad\qquad+\left.\left.\int_{]0,T]}\beta_u^{-1}L_{u-}\Bigl(
Z^1(u,S_{u-})dH_u^1+Z^2(u,S_{u-})dH_u^2\Bigr)\right|\calf_t\right]=m_t
\eea
where 
\be
q(t,v)=y_t^1\delta_t^1\bold{1}_{\{v<0\}}+y_t^2\delta_t^2\bold{1}_{\{v\geq 0\}}
\ee
and $\{m_t\}_{t\geq 0}$ is a $Q$-martingale. Thus we obtain the following SDE:
\be
dS_t-r_tS_tdt + L_t\bigl(dD_t+q(t,S_t)S_tdt\bigr)
+L_{t-}\bigl(Z^1(t,S_{t-})dH_t^1+Z^2(t,S_{t-})dH_t^2\bigr)=\beta_t dm_t~.
\ee
Using the decomposition of $H_t^i$, we get
\be
dS_t-r_tS_tdt+L_t\bigl(dD_t+q(t,S_t)S_tdt\bigr)
+L_t\bigl(Z^1(t,S_t)h_t^1 + Z^2(t,S_t)h_t^2\bigr)dt=dn_t~,
\ee
where we have defined
\be
dn_t=\beta_t dm_t-L_{t-}\bigl(Z^1(t,S_{t-})dM_t^1 +Z^2(t,S_{t-})dM_t^2\bigr)
\ee
and $\{n_t\}_{t\geq 0}$ is also a some $Q$-martingale.
Using the fact that
\be
q(t,S_t)S_t+Z^1(t,S_t)h_t^1+Z^2(t,S_t)h_t^2=S_t\bigl(\mu(t,S_t)+h_t\bigr)~,
\ee
one can show that the SDE for $S_t$ is given by
\be
dS_t=-L_t dD_t+L_t\bigl(r_t-\mu(t,S_t)-h_t\bigr)S_tdt+dn_t~.
\label{S_SDE}
\ee

Secondly, let us consider the SDE for $V_t$. By following the similar procedures,
one can easily see that
\bea
&&e^{-\int_0^t \bigl(r_u-\mu(u,V_u)\bigr)du}V_t+\int_{]0,t]}e^{-\int_0^s\bigl(r_u-\mu(u,V_u)\bigr)du}dD_s\nonumber\\
&&=E^Q\left[\left.\int_{]0,T]}\exp\left(-\int_0^s \bigl(r_u-\mu(u,V_u)\bigr)du\right)dD_u\right|\calf_t\right]=\tilde{m}_t~,
\eea
where $\{\tilde{m}_t\}_{t\geq 0}$ is a $Q$-martingale.
Thus we have
\be
dV_t=-dD_t+\bigl(r_t-\mu(t,V_t)\bigr)V_tdt+d\tilde{n}_t~,
\ee
where
\be
d\tilde{n}_t=e^{\int_0^t \bigl(r_u-\mu(u,V_u)\bigr)du}d\tilde{m}_t~,
\ee
and hence $\{\tilde{n}_t\}_{t\geq 0}$ is also a Q-martingale.
As a result we have
\bea
d(\bold{1}_{\{\tau>t\}}V_t)&=&d(L_t V_t)\nonumber\\
&=&L_{t-}dV_t-V_{t-}dH_t-\Delta V_\tau \Delta H_{\tau}\nonumber\\
&=&-L_{t-}dD_t+L_{t}\bigl(r_t-\mu(t,V_t)\bigr)V_tdt-L_tV_th_tdt-\Delta V_\tau \Delta H_{\tau}\nonumber\\
&&+L_{t-}\left(d\tilde{n}_t-V_{t-}(dM_t^1+dM_t^2)\right)\nonumber\\
&=&-L_tdD_t+L_t\bigl(r_t-\mu(t,V_t)-h_t\bigr)V_tdt-\Delta V_\tau \Delta H_{\tau}+d\tilde{N}_t,
\label{V_SDE}
\eea
where $\{\tilde{N}_t\}_{t\geq 0}$ is a Q-martingale such that
\be
d\tilde{N}_t=L_{t-}\left(d\tilde{n}_t-V_{t-}(dM_t^1+dM_t^2)\right)~.
\ee
Therefore, by comparing Eqs.~(\ref{S_SDE}) and (\ref{V_SDE}) and also the fact that
$S_T=\bold{1}_{\{\tau>T\}}V_T=0$, we cannot distinguish $\bold{1}_{\{\tau>t\}}V_t$ from $S_t$ if
there is no jump at the time of default $\Delta V_\tau=0$.~$\blacksquare$
\\
\\
{\it Remark}:~ In this remark, we briefly discuss the assumption of $\Delta V_{\tau}=0$.
Notice that, since we assume totally inaccessible default time, there is no contribution from pre-fixed lump-sum coupon payments
to the jump. In addition, it is natural (and also common in the existing literatures) to assume global market variables, such as interest rates and FX's, are adapted to the background filtration independent from the defaults.
In this paper, we are concentrating on the standard fixed income derivatives without credit sensitive dividends, and 
hence the only thing we need to care about is the behavior of hazard rates, $h^1$ and $h^2$.
Therefore, in this case, if there is no jump on $h^i$ on the default of the other party 
$j\neq i$, then the assumption $\Delta V_{\tau}=0$ holds true.
This corresponds to the situation where there is no default dependence between the two firms.

If there exists non-zero default dependence, which is important in risk-management point of view, 
then there appears a jump on the hazard rate of the surviving firm when a default occurs.
This represents a direct feedback (or a contagious effect) from the defaulted firm to the 
surviving one.
In this case, if we directly use $\mathbb{F}$-intensities $h^i$, the 
no-jump assumption does not hold. 

However, even in this case, there is a way to handle the pricing problem 
correctly.
Let us construct the filtration in the usual 
way as $\calf_t=\calg_t\vee\calh^1_t\vee\calh^2_t$, where $\calg_t$ is the background filtration (say, generated by Brownian motions), and 
$\calh_t^i$ is the filtration generated by $H^i$. Since the only information we need is up to $\tau=\tau^1\wedge\tau^2$,
we can limit our attention to the intensities conditional on no-default, which are now the processes adapted to the 
background filtration $\mathbb{G}=(\calg_t)_{\{t\geq 0\}}$. 
Therefore, although the details of the derivation slightly change, one can show that the pricing 
formula given in Eq.~(\ref{eqV}) can still be applied in the same way once we use
the $\mathbb{G}$-intensities instead, since now we can write all the processes involved in the formula adapted to the 
background filtration. 

%%%%%%%%%%%%%%%%%%%%%%%%%%%%%%%%%%%%%%%%%%%%%%%%%%%%%%%%%%%%%%%%%%%%%%%%%
\section{Origin of the Funding Spread $y^{(i,k)}$ in the Pricing Formula}
\label{funding_spread}
%%%%%%%%%%%%%%%%%%%%%%%%%%%%%%%%%%%%%%%%%%%%%%%%%%%%%%%%%%%%%%%%%%%%%%%%
Here, let us comment about the origin of the funding spread $y$ in our pricing formula in Eq.~(\ref{perfect_foreign}).
Consider the following hypothetical but plausible situation to get a clear image:
\\ \\
{\it(1): An interest rate swap market where the participants are discounting future cash flows by
domestic OIS rate, regardless of the collateral currency, and assume there is no 
price dispute among them.
(2): Party $1$ enters two opposite trades with party $2$ and $3$, and they are 
agree to have CSA which forces party $2$ and $3$ to always post a domestic currency $U$ 
as collateral, but party $1$ is allowed to use a foreign currency $E$  as well as $U$. 
(3): There is very liquid CCOIS market which allows firms to enter arbitrary length of swap. The spread $y$
is negative for  CCOIS  between $U$ and $E$, where $U$ is a base
currency (such as USD in the above explanation).} 
\\\\
In this example, the party $1$ can definitely make money.
Suppose, at a certain point, the party $1$ receives $N$ unit amount of $U$ from the party $2$ as collateral.
Party $1$ enters a CCOIS as spread payer, 
exchanging  $N$ unit amount of $U$ and the corresponding amount of $E$,
by which it can finance the foreign currency $E$ by the rate of ($E$'s OIS $+y^{(E,U)}$).
Party $1$ also receives $U$'s OIS rate from the CCOIS counter party, which is going to be paid
as the collateral margin to the party $2$.
Party $1$ also posts $E$ to the party $3$ since it has opposite position, 
it receives $E$'s OIS rate as the collateral margin from 
the party $3$.  As a result, the party $1$ earns $-y^{(E,U)}~(>0)$ on the notional amount of collateral.
It can rollover the CCOIS, or unwind it if $y$'s sign flips.

Of course, in the real world, CCS can only be traded with certain terms which makes
the issue not so simple. However, considering significant size of CCS spread (a several tens of bps)
it still seems possible to arrange appropriate CCS contracts to achieve cheaper funding.
For a very short term, it may be easier to use FX forward contracts (or FX swaps) for the same purpose. 
In order to prohibit this type of arbitrage, party $1$ should pay extra premium 
to make advantageous CSA contracts. This is exactly the reason why our pricing formula 
contains the funding spread $y$.

%%%%%%%%%%%%%%%%%%%%%%%%%%%%%%%%%%%%%%%%%%%%%%%%%%%%%%%%%%%%%%%%%%%%%%
\section{Calibration to swap markets}
\label{curve_construction}
%%%%%%%%%%%%%%%%%%%%%%%%%%%%%%%%%%%%%%%%%%%%%%%%%%%%%%%%%%%%%%%%%%%%%%
For the details of calibration procedures, the numerical results and 
recent historical behavior of underlyings are available in Refs.~\cite{multiple_curves, collateral_choice}. 
The procedures can be briefly summarized as follows:
(1) Calibrate the forward collateral rate $c^{(i)}(0,t)$ for each currency using OIS market.
(2) Calibrate the forward Libor curves by using the result of (1), IRS and tenor swap markets.
(3) Calibrate the forward $y^{(i,j)}(0,t)$ spread for each relevant currency pair 
by using the results of (1),(2) and CCS markets.

Although we can directly obtains the set of $y^{(i,j)}$ from CCS, we cannot uniquely determine
each $y^{(i)}$, which is necessary for the evaluation of Gateaux derivative when we deal with 
unilateral collateralization and CCA (collateral cost adjustment). For these cases, we need to make an assumption 
on the risk-free rate for one {\it and only one} currency. For example, if we assume that ON rate and 
the risk-free rate of currency $(j)$ are the same, and hence $y^{(j)}=0$,  then the forward curve of $y^{\rm USD}$ is 
fixed by $y^{\rm USD}(0,t)=-y^{(j,\rm{USD})}(0,t)$. Then using the result of $y^{\rm USD}$,
we obtains $\{y^{(k)}\}$ for all the other currencies by making use of $\{y^{(k,\rm USD)}\}$ obtained from CCS
markets. More ideally, each financial firm may carry out some analysis on 
the risk-free profit rate of cash pool or more advanced econometric analysis on the 
risk-free rate, such as those given in Feldh\"utter \& Lando (2008)~\cite{Lando}.

%%%%%%%%%%%%%%%%%%%%%%%%%%%%%%%%%%%%%%%%%%%%%%%%%%%%%%%%%%%%%
\section{Details of Present Value Derivation in Sec.~\ref{numerical}}
%%%%%%%%%%%%%%%%%%%%%%%%%%%%%%%%%%%%%%%%%%%%%%%%%%%%%%%%%%%%%
\subsection{Asymmetrically collateralized MtMCCOIS}
\label{asymmetric_MtMCCOIS}
%%%%%%%%%%%%%%%%%%%%%%%%%%%%%%%%%%%%%%%%%%%%%%%%%%%%%%%%%%%%%%
In this case, the price of the contract at time $0$ from the view point of party $1$ is given by
\be
V_0=E^{Q^{(j)}}\left[\int_{]0,T_N]} \exp\left(-\int_0^s R(u,V_u) du\right)dD_s\right]
\ee
where 
\be
R(t,V_t)=c^{(j)}_t+y^{(j,i)}_t+\max\bigl(-y^{(j,i)}_t,0\bigr)\bold{1}_{\{V_t<0\}}~,
\ee
and
\be
dD_s=\sum_{n=1}^N\left\{\delta_{T_n}(s)
\left[-e^{\int_{T_{n-1}}^{T_n}c^{(j)}_udu}-\delta_n B+\frac{f_x^{(j,i)}(T_n)}{f_x^{(j,i)}(T_{n-1})}
e^{\int_{T_{n-1}}^{T_n}c^{(i)}_udu}\right]\right\}~.
\ee
The expression of $R$ can be easily checked by noticing that
\bea
y^1&=&\min(y^{(i)},y^{(j)})=y^{(i)}+\min(y^{(j,i)},0)\nn \\
&=&y^{(i)}-\max(-y^{(j,i)},0) \\
y^2&=&y^{(i)}
\eea

Using Gateaux derivative, we can approximate the contract price as
\be
V_0\simeq \overline{V}_0+{\rm CCA}_0
\ee
where 
\bea
&&{\rm CCA}_0= E^{Q^{(j)}}\left[
\int_0^{T_N} e^{-\int_0^s (c_u^{(j)}+y_u^{(j,i)})du}\max\bigl(-\overline{V}_s,0\bigr)\max\bigl(-y_s^{(j,i)},0
\bigr)ds\right]~.
\label{cca_ccs}
\eea
Although $\overline{V}_t$ is simply a price under symmetric collateralization using currency $(i)$, 
we need to be careful about the advance reset conventions.
One can show that 
\bea
\overline{V}_t&=&\sum_{n=\gamma(t)+1}^N E^{Q^{(j)}}\left[
\left. e^{-\int_t^{T_n}(c_u^{(j)}+y_u^{(j,i)})du}\left\{-e^{\int_{T_{n-1}}^{T_n}c_u^{(j)}du}-\delta_n B
+\frac{f_x^{(j,i)}(T_n)}{f_x^{(j,i)}(T_{n-1})}e^{\int_{T_{n-1}}^{T_n}c_u^{(i)}du}\right\}\right|\calf_t\right]\nonumber\\
&&\hspace{-28mm}+E^{Q^{(j)}}\left[\left.
e^{-\int_t^{T_{\gamma(t)}}(c^{(j)}_u+y_u^{(j,i)})du}\left\{ -e^{\left(\int_{T_{\gamma(t)-1}}^{t}+\int_t^{T_{\gamma(t)}}\right)
c^{(j)}_u du}-\delta_{\gamma(t)} B +\frac{e^{\int_{T_{\gamma(t)-1}}^t c^{(i)}_udu}}{f_x^{(j,i)}(T_{\gamma(t)-1})}
e^{\int_t^{T_{\gamma(t)}}c^{(i)}_udu}f_x^{(j,i)}(T_{\gamma(t)}) \right\}
\right| \calf_t\right]~,\nonumber\\
\eea
where $\gamma(t)=\min\{n; T_n>t, n=1\cdots N\}$.
Note that $T_{\gamma(t)-1}<t$ since we are considering spot-start swap (or $T_0=0$). Assuming the 
independence of $y^{(j,i)}$ and other variables, we can simplify $V_t(0)$ and obtains
\bea
\overline{V}_t&=&-\sum_{n=\gamma(t)}^ND^{(j)}(t,T_n)Y^{(j,i)}(t,T_n)\delta_n B
+\sum_{n=\gamma(t)+1}^N D^{(j)}(t,T_{n-1})\left(Y^{(j,i)}(t,T_{n-1})-Y^{(j,i)}(t,T_n)\right)\nonumber\\
&&-Y^{(j,i)}(t,T_{\gamma(t)})e^{\int_{T_{\gamma(t)-1}}^t c^{(j)}_s ds}
+\frac{f_x^{(j,i)}(t)}{f_x^{(j,i)}(T_{\gamma(t)-1})}e^{\int_{T_{\gamma(t)-1}}^t c^{(i)}_sds}~,
\eea
where we have defined $ Y^{(j,i)}(t,T)=E^{Q^{(j)}}\left[\left.e^{-\int_t^T y_s^{(j,i)}ds}\right|\calf_t\right]$.
We need to evaluate the above $\overline{V}$ at each time step of Monte Carlo simulation to 
calculate {\rm CCA} of Eq.~(\ref{cca_ccs}).

%%%%%%%%%%%%%%%%%%%%%%%%%%%%%%%%%%%%%%%%%%%%%%%
\subsection{Asymmetrically collateralized OIS}
\label{asymmetric_OIS}
%%%%%%%%%%%%%%%%%%%%%%%%%%%%%%%%%%%%%%%%%%%%%%%%
For spot-start, $T_N$-maturing OIS, we have 
\be
V_0=E^{Q^{(j)}}\left[\int_{]0,T_N]}e^{-\int_0^s R(u,V_u)du}dD_s\right]~,
\ee
where
\be
dD_s=\sum_{n=1}^N\delta_{T_n}(s)\left[\delta_n S-\left(e^{\int_{T_{n-1}}^{T_n}c^{(j)}_udu}-1\right)\right]~,
\ee
and
\be
R(t,V_t)=c^{(j)}_t +\max(y^{(j,i)}_t,0)\bold{1}_{\{V_t<0\}}~.
\ee
Using Gateaux derivative, the above swap value can be approximated as
\be
V_0\simeq \overline{V}_0+{\rm CCA}_0,
\ee
where
\be
{\rm CCA}_0=
E^{Q^{(j)}}\left[\int_0^T e^{-\int_0^s c^{(j)}_udu}\max\bigl(-\overline{V}_s,0\bigr)\max\bigl(y^{(j,i)}_s,0\bigr)ds\right],
\ee
and 
\bea
\overline{V}_t&=&E^{Q^{(j)}}\left[\left. \sum_{n=\gamma(t)}^N e^{-\int_t^{T_n}c^{(j)}_u du}
\left\{\delta_n S-\left(e^{\int_{T_{n-1}}^{T_n}c^{(j)}_udu}-1\right)\right\}\right|\calf_t\right]\nonumber\\
&=&\sum_{n=\gamma(t)}^ND^{(j)}(t,T_n)\delta_n S-e^{\int_{T_{\gamma(t)-1}}^t c^{(j)}_udu}+D^{(j)}(t,T_N)~.
\eea
Here, $S$ is the fixed OIS rate.

%%%%%%%%%%%%%%%%%%%%%%%%%%%%%%%%%%%%%%%%%%%%%%%%%%%%%%%%%%
\section{ Proof of Proposition~\ref{prop-2}}\label{proof-2}
%%%%%%%%%%%%%%%%%%%%%%%%%%%%%%%%%%%%%%%%%%%%%%%%%%%%%%%%%
Consider the case of $y^1\geq y^2$.
From Eq.~(\ref{eqV}), one can show that the pre-default value $V$ can also be written 
in the following recursive form:
\be
V_t=E^{Q}\left[\left.
-\int_{]t,T]}\bigl(r_s-\mu(s,V_s)\bigr)V_sds+\int_{]t,T]}dD_s\right|\calf_t\right]~.
\label{V_recursive}
\ee
Let us define the following variables:
\bea
\tilde{V}_t&=&e^{-\int_0^t (r_s-y_s^1)ds}V_t\\
\tilde{D}_t&=&\int_{]0,t]}e^{-\int_0^s (r_u-y_u^1)du}dD_s~.
\eea
Note that
\bea
r_t-\mu(t,V_t)&=&(r_t-y_t^1)+(y_t^1-y_t^2)\bold{1}_{\{V_t\geq 0\}}~\nonumber\\
&=&(r_t-y_t^1)+\eta_t^{1,2}\bold{1}_{\{V_t\geq 0\}}~,
\eea
where we have defined $\eta^{i,j}=y^i-y^j$.
Using new variables, Eq.~(\ref{V_recursive}) can be rewritten as
\be
\tilde{V}_t=E^{Q}\left[\left. -\int_{]t,T]}\eta_s^{1,2}\bold{1}_{\{\tilde{V}_s\geq 0\}}\tilde{V}_s ds
+\int_{]t,T]}d\tilde{D}_s\right|\calf_t\right]~.
\ee
And hence we have,
\be
\tilde{V}_t^{ab}-\tilde{V}_t^a-\tilde{V}_t^b=E^{Q}\left[\left.
-\int_{]t,T]}\eta_s^{1,2}\left(
\max\bigl(\tilde{V}_s^{ab},0\bigr)-\max\bigl(\tilde{V}_s^a,0\bigr)-\max\bigl(\tilde{V}^b_s,0\bigr)\right)ds\right| 
\calf_t\right]~.
\ee
Let us denote the upper bound of $\eta^{1,2}$ as $\alpha$, and also define $Y=\tilde{V}^{ab}-\tilde{V}^a-\tilde{V}^b$
and $G_s=-\eta_s^{1,2}\left(
\max\bigl(\tilde{V}_s^{ab},0\bigr)-\max\bigl(\tilde{V}_s^a,0\bigr)-\max\bigl(\tilde{V}^b_s,0\bigr)\right)$.
Then, we have $Y_T=0$ and 
\be
Y=E^{Q}\left[\left.\int_{]t,T]}G_sds\right|\calf_t\right].
\ee

\bea
G_s&=&-\eta_s^{1,2}\left(\max\bigl(\tilde{V}_s^{ab},0\bigr)-\max\bigl(\tilde{V}_s^a,0\bigr)-\max\bigl(\tilde{V}^b_s,0\bigr)\right)\nonumber\\
&\geq&-\eta_s^{1,2}\left(\max(\tilde{V}_s^{ab},0)-\max(\tilde{V}_s^a+\tilde{V}_s^b,0)\right)\nonumber\\
&\geq&-\eta_s^{1,2}\max\left(\tilde{V}_s^{ab}-\tilde{V}^a_s-\tilde{V}^b_s,0\right)\nonumber\\
&\geq&-\alpha |Y_s|~.
\eea
Applying the consequence of the Stochastic Gronwall-Bellman Inequality in Lemma $\bold{B2}$ of Ref.~\cite{Duffie-Epstein}
to $Y$ and $G$, we can conclude $Y_t\geq 0$ for all $t\in [0,T]$,
and hence $V^{ab}\geq V^a+V^b$.$\blacksquare$

%%%%%%%%%%%%%%%%%%%%%%%%%%%%%%%%%%%%%%%%%%%%%%%%%%%%%%%%%%%%%%
\section{ Proof of Proposition~\ref{prop-3}}\label{proof-3}
%%%%%%%%%%%%%%%%%%%%%%%%%%%%%%%%%%%%%%%%%%%%%%%%%%%%%%%%%%%%%%
Consider the case of $y^1\geq y^2$. Let us define
\bea
\tilde{V}^{F}_t&=&e^{-\int_0^t (r_s-y_s^1)ds}V_t^{F}\\
\tilde{V}^{G}_t&=&e^{-\int_0^t (r_s-y_s^1)ds}V_t^{G}~,
\eea
as well as
\be
\tilde{D}_t =\int_{]0,t]}e^{-\int_0^s (r_u-y_u^1)du}dD_s
\ee
as in the previous section. Then, we have
\bea
\tilde{V}^G_t&=&E^{Q}\left[\left. -\int_{]t,T]}\eta^{1,2}_s\max(\tilde{V}_s^{G},0)ds+
\int_{]t,T]}d\tilde{D}_s\right|\calg_t\right]\\
\tilde{V}^F_t&=&E^{Q}\left[\left. -\int_{]t,T]}\eta^{1,2}_s\max(\tilde{V}_s^{F},0)ds+
\int_{]t,T]}d\tilde{D}_s\right|\calf_t\right]~.
\eea
Now, let us define 
\be
U_t=E^Q\left[\left.\tilde{V}_t^{G}\right|\calf_t\right]~.
\ee
Then, using Jensen's inequality, we have
\be
U_t\leq E^Q\left[\left. -\int_{]t,T]}\eta_s^{1,2}\max(U_s,0)ds+\int_{]t,T]}d\tilde{D}_s\right|\calf_t\right]~.
\ee
Therefore, we obtain
\bea
\tilde{V}_t^F-U_t &\geq& E^Q\left[\left. -\int_{]t,T]}\eta^{1,2}_s
\left(\max(\tilde{V}_s^F,0)-\max(U_s,0)\right)ds\right|\calf_t\right]\\
&\geq &E^Q\left[\left. -\int_{]t,T]}\eta_s^{1,2}\bigl|\tilde{V}_s^F-U_s\bigr|ds\right|\calf_t\right]~.
\eea
Using the stochastic Gronwall-Bellman Inequality as before, one can conclude that
$\tilde{V}_t^F\geq U_t$ for all $t\in[0,T]$, and in particular, $V_0^F\geq V_0^G$.$\blacksquare$

%%%%%%%%%%%%%%%%%%%%%%%%%%%%%%%%%%%%%%%%%%%%%%%%%%%%%%%%%%%%%
\section{Comparison of Gateaux Derivative with PDE}\label{PDE}
%%%%%%%%%%%%%%%%%%%%%%%%%%%%%%%%%%%%%%%%%%%%%%%%%%%%%%%%%%%%%
In order to get clear image for the reliability of Gateaux derivative, we compare it 
with the numerical result directly obtained from PDE.
We consider a simplified setup where
MtMCCOIS exchanges the coupons continuously, and the only stochastic variable is a spread $y$. 
Consider continuous payment $(i,j)$-MtMCCOIS where the leg of currency $(i)$ needs notional 
refreshments. We assume following situation as the asymmetric collateralization:\\
(1) Party $1$ is the basis spread payer and can use either the currency $(i)$ or $(j)$ as collateral.\\
(2) Party $2$ is the basis spread receiver and can only use the currency $(i)$ as collateral. \\

In this case, one can see that the value of $t$-start $T$-maturing contract from the view point of party $1$ is
given by (See, Eq.~(\ref{CCOIS_k}).)
\be
V_t=E^{Q^{(j)}}\left[\left. \int_t^T \exp\left(-\int_t^s R(u,V_u)du\right)\left(
y^{(j,i)}_s-B\right)ds\right|\calf_t\right]~,
\ee
where
\be
R(t,V_t)=c^{(j)}(t)+y^{(j,i)}_t+\max\left(-y^{(j,i)}_t,0\right)\bold{1}_{\{V_t<0\}}~
\ee
and $B$ is a fixed spread for the contract. $y^{(j,i)}$ is the only stochastic variable and its 
dynamics is assumed to be given by the following Hull-White model:
\be
dy^{(j,i)}_t=\left(\theta^{(j,i)}(t)-\kappa^{(j,i)} y^{(j,i)}_t\right)dt +\sigma_y^{(j,i)} dW_t^{Q^{(j)}}.
\ee
Here, $\theta^{(j,i)}(t)$ is a deterministic function specified by the initial term structure of $y^{(j,i)}$,
$\kappa^{(j,i)}$ and $\sigma_y^{(j,i)}$ are constants. $W^{Q^{(j)}}$ is a Brownian motion under the spot martingale 
measure of currency $(j)$.

The PDE for $V_t$ is given by
\bea
\frac{\partial}{\partial t}V(t,y)+\left(\gamma(t,y)\frac{\partial V(t,y)}{\partial y}
+\frac{\bigl(\sigma_y^{(j,i)}\bigr)^2}{2}\frac{\partial^2}{\partial y^2}V(t,y)\right)-R\bigl(t,V(t,y)\bigr)V(t,y)
+y-B=0~, \nonumber\\
\eea
where
\be
\gamma(t,y)=\theta^{(j,i)}(t)-\kappa^{(j,i)}y~.
\ee
If party $1$ is a spread receiver, we need to change $y-B$ to $B-y$, of course.

Terminal boundary condition is trivially given by $V(T,\cdot)=0$.
On the lower boundary of $y$ or when $y=-M~(=y_{\min})\ll 0$, 
we have $V_t<0$ for all $t$. Thus, we have $R(s,V(s,y))=c^{(j)}(s)$ for all $s\geq t$, if 
$y=-M$ at time $t$.
Therefore, on the lower boundary, the value of MtMCCOIS is given by
\bea
V(t,-M)&=&E^{Q^{(j)}}\left[\left. \int_t^T e^{-\int_t^s c_u^{(j)}du}(y^{(j,i)}_s-B)ds\right|y^{(j,i)}_t=-M\right]\nonumber\\
&=&\int_t^T D^{(j)}(t,s)\left(-B-\frac{\partial}{\partial s}\ln Y^{(j,i)}(t,s)\right)ds~.
\eea
%where we have defined
%\be
%Y^{(j,i)}(t,s)=E^{Q^{(j)}}\left[\left.e^{-\int_t^s y^{(j,i)}_u du}\right|\calf_t\right]~,
%\ee
%which can be expressed by $y_t=-M$ and the initial term structure of $y$ at time $0$ due to its
%affine property. 
Since $c^{(j)}(t)$ is a deterministic function, $D^{(j)}(t,s)=D^{(j)}(0,s)/D^{(j)}(0,t)$
is simply given by the forward.

On the other hand, when $y=M~(=y_{\max})\gg 0$, we have $V_t>0$ for all $t$. 
Thus we have $R(s,V(s,y))=c^{(j)}(s)+y^{(j,i)}(s)$ for all $s\geq t$, if $y=M$ at time $t$.
Thus, on the upper boundary, the value of the contract becomes
\bea
V(t,M)&=&E^{Q^{(j)}}\left[\left. \int_t^T e^{-\int_t^s \bigl(c^{(j)}_u+y_u^{(j,i)}\bigr)du}
\left(y^{(j,i)}_s-B\right)\right|y_t^{(j,i)}=M\right]\nonumber\\
&=&\int_t^T\left\{ -BD^{(j)}(t,s)Y^{(j,i)}(t,s)-D^{(j)}(t,s)\frac{\partial}{\partial s}Y^{(j,i)}(t,s)
\right\}ds~.
\eea

Now let us compare the numerical result between Gateaux derivative and PDE.
In the case of Gateaux derivative, the contract value is approximated as
\be
V_t\simeq \overline{V}_t+\nabla V_t,
\ee
where
\be
\overline{V}_t=E^{Q^{(j)}}\left[\left. \int_t^T e^{-\int_t^s (c^{(j)}_u+y^{(j,i)}_u)du}\left(y^{(j,i)}_s-B\right)ds
\right|\calf_t\right]~,
\ee
and 
\bea
\nabla V_t= E^{Q^{(j)}}\left[\left.\int_t^T e^{-\int_t^s (c_u^{(j)}+y^{(j,i)}_u)du}\bigl[-\overline{V}_s\bigr]^+
\max\bigl(-y^{(j,i)}_s,0\bigr)ds\right|\calf_t\right].
\eea
$\overline{V}_t$ is the value of the contract under symmetric collateralization where both parties post currency $(i)$ as
collateral, and $\nabla V_t$ is a deviation from it.

\begin{figure}
	\center{\includegraphics[width=120mm]{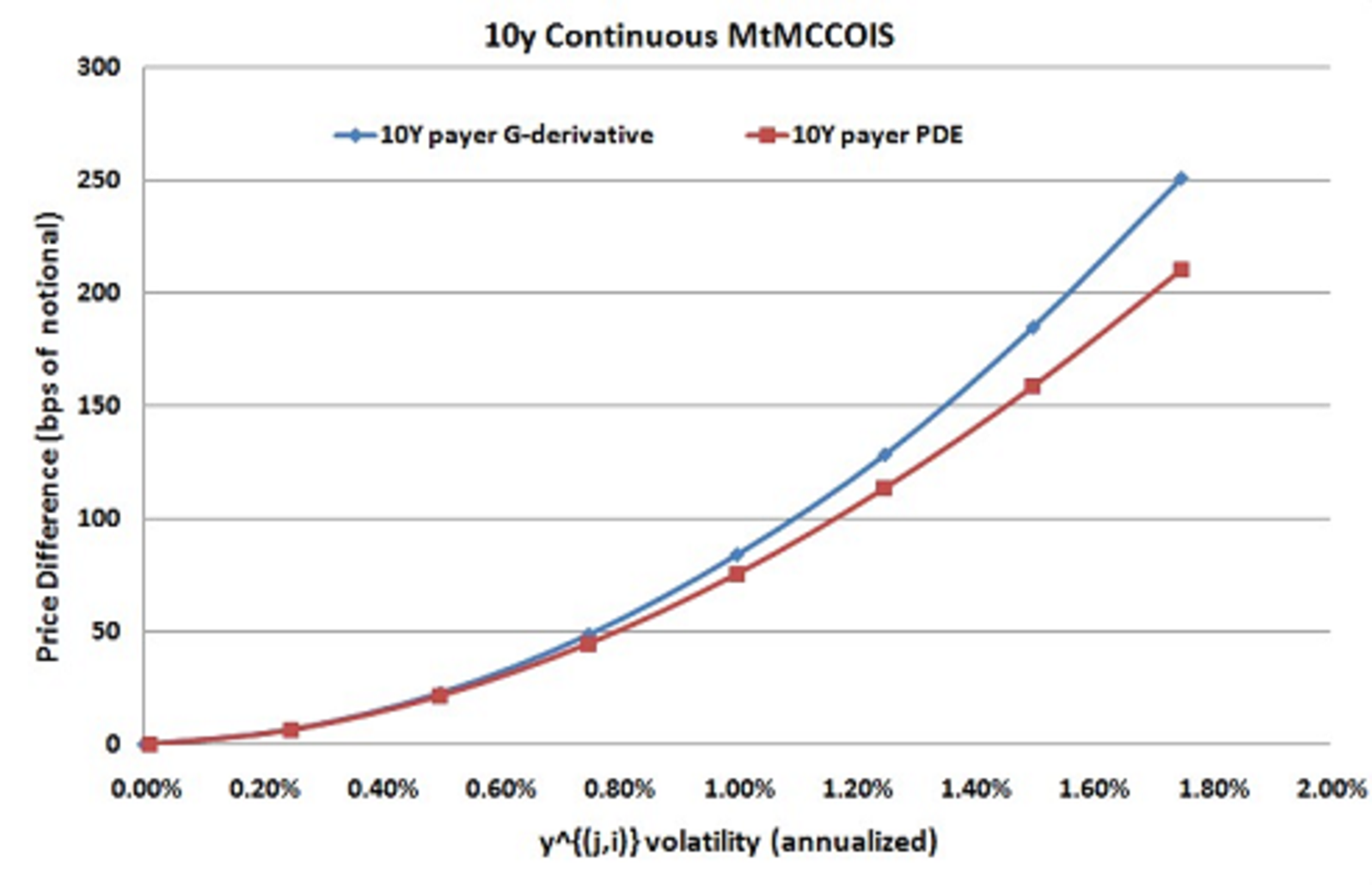}}
	\vspace{-5mm}
	\caption{Price difference from symmetric limit for 10y continuous MtMCCOIS}
	\label{continuousCCOIS}
\end{figure}

In Fig.~\ref{continuousCCOIS}, we plot the price difference of continuous 10y-MtMCCOIS from 
its symmetric limit obtained by PDE and 
Gateaux derivative with various volatility of $y^{(j,i)}$.  Term structures of $y^{(j,i)}$ and other curves 
are given in Appendix~\ref{Data}. Here, the spread $B$ is chosen in such a way that 
the swap price is zero in the case where both parties can only use currency $(i)$ as collateral, or B is a market par spread.
The price difference is $V_t-\overline{V}_t$ and expressed as basis points of notional.
From our analysis using the recent historical data in Ref.~\cite{collateral_choice},
we know that the annualized volatility of $y$ is around $50$ bps for a calm market but it can be more than $(100\sim 200)$ bps
when CCS market is volatile (We have used EUR/USD and USD/JPY pairs.). 
One observes that Gateaux derivative provides reasonable approximation for wide range of volatility.
If the party $1$ is a spread receiver, both of the methods give very small price differences, less than $1$bp of notional.

%%%%%%%%%%%%%%%%%%%%%%%%%%%%%%%%%%%%%%%%%%%%%%%%%%%%%%%%%%%%%%%
\section{Data used in  Numerical Studies}\label{Data}
%%%%%%%%%%%%%%%%%%%%%%%%%%%%%%%%%%%%%%%%%%%%%%%%%%%%%%%%%%%%%%%%
The parameter we have used in simulation are
\bea
&&\kappa^{(j)}=\kappa^{(i)}=1.5\% \\
&&\sigma_c^{(j)}=\sigma_c^{(i)}=1\% \\
&&\sigma_x^{(j,i)}=12\% ~.
\eea
All of them are defined in annualized term.
The volatility of $y^{(j,i)}$ is specified in the main text in each numerical analysis.

\begin{figure}[h]
	\center{\includegraphics[width=123mm]{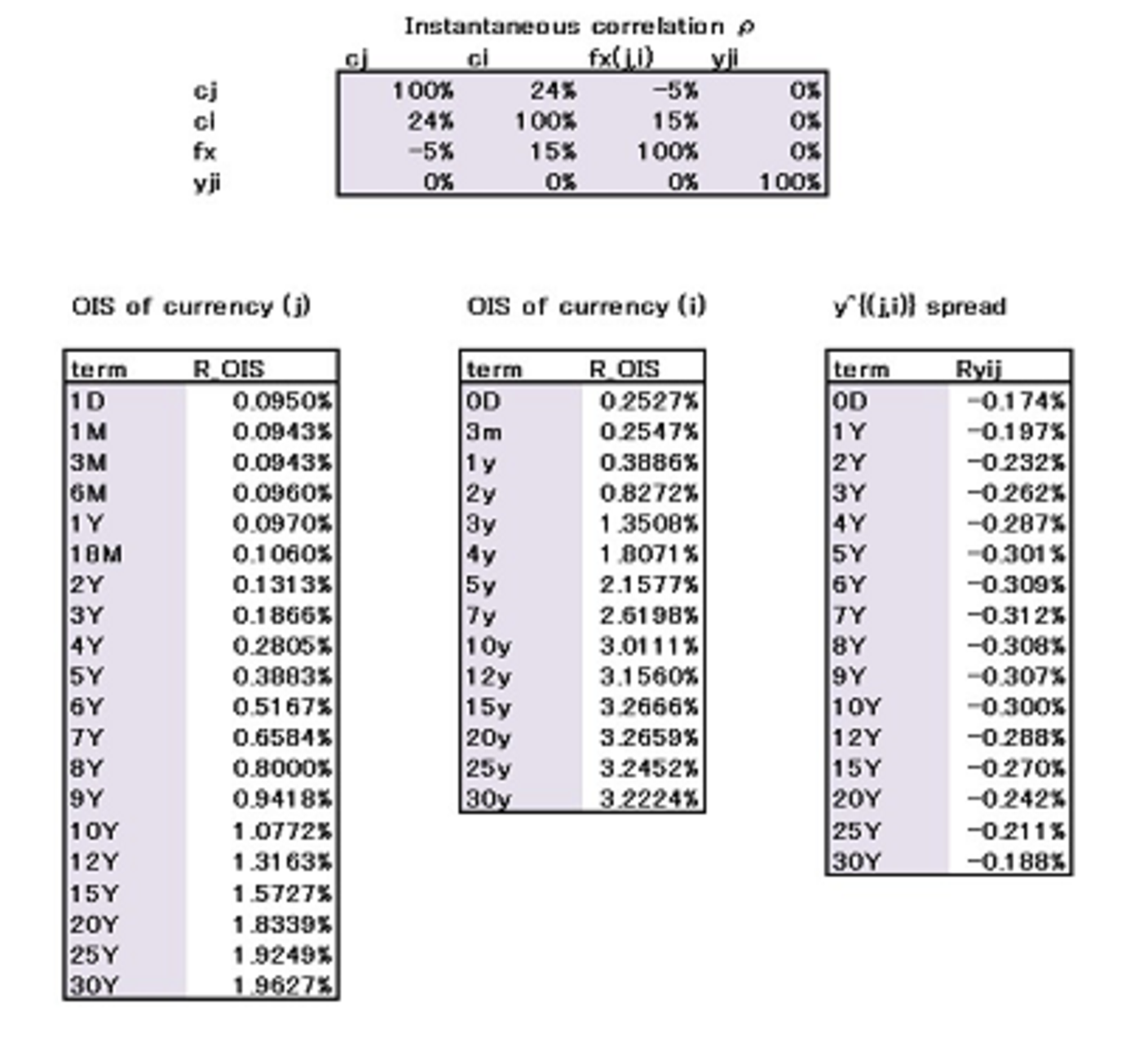}}
	\caption{Term structures and correlation used for simulation}
	\label{simData}
\end{figure}

Term structures and correlation used in simulation are given in Fig.~\ref{simData}.
There we have defined
\bea
R^{(k)}_{\rm OIS}(T)&=&-\frac{1}{T}\ln E^{Q^{(k)}}\left[ e^{-\int_0^T c^{(k)}_s ds}\right] \nonumber\\
R_{y^{(j,i)}}(T)&=&-\frac{1}{T}\ln E^{Q^{(j)}}\left[ e^{-\int_0^T y^{(j,i)}_s ds}\right]~. \nonumber
\eea
The curve data is based on the calibration result of typical JPY and USD market data of early 2010.
In Monte Carlo simulation, in order to reduce simulation error, we have adjusted drift terms $\theta(t)$ to achieve exact 
match to the relevant forwards in each time step.

%%%%%%%%%%%%%%%%%%%%%%%%%%%%%%%%%%%%%%%%%%%%%%%%%%%%%%%%%%%%%%%%%%%%%%%%%%%%%

%%%%%%%%%%%%%%%%%%%%%%%%%%%%%%%%%%%%%%%%%%%%%%%%%%%%%%%%%%%%%%%%%%%%
\end{document}